\documentclass[%
 aip,
 amsmath,amssymb,
 preprint,%
]{revtex4-1}

\usepackage{amsmath, amssymb, accents, float} % note that this was added
\usepackage{makecell}

\usepackage{graphicx}% Include figure files
\usepackage{dcolumn}% Align table columns on decimal point
\usepackage{bm}% bold math
\usepackage[utf8]{inputenc}
\usepackage[T1]{fontenc}
\usepackage{mathptmx}
\usepackage{etoolbox}

\makeatletter
\def\@email#1#2{%
 \endgroup
 \patchcmd{\titleblock@produce}
  {\frontmatter@RRAPformat}
  {\frontmatter@RRAPformat{\produce@RRAP{*#1\href{mailto:#2}{#2}}}\frontmatter@RRAPformat}
  {}{}
}%
\makeatother
\begin{document}

\preprint{AIP/123-QED}

\title[Exp Shear]{A new approach to measuring both the steady flow curve and the planar extensional viscosity of complex fluids using exponential shear}
% Force line breaks with \\
\author{Nicholas King}
\affiliation{ 
Department of Chemical Engineering, Massachusetts Institute of Technology%\\This line break forced with \textbackslash\textbackslash
}%

\author{Eugene Pashkovski}%
\author{Reid Patterson}
\affiliation{ 
Lubrizol Corporation, Research and Development%\\This line break forced with \textbackslash\textbackslash
}%

\author{Gareth H. McKinley *}
\email{gareth@mit.edu}
\affiliation{%
Department of Mechanical Engineering, Massachusetts Institute of Technology%\\This line break forced% with \\
}%

\date{\today}% It is always \today, today,

\begin{abstract}
We revisit the protocol known as Exponential Shear (ES), first proposed by Doshi and Dealy in 1986. This is a transient strong flow that can be readily imposed in the laboratory using commercial strain-controlled (separate motor and transducer) rheometers which accurately measure both the (exponentially) time-varying torque and normal force exerted by the sample. Numerous studies over the years have proposed competing definitions of an appropriate material function for reporting data obtained from the exponentially time-varying imposed deformation. We revisit this unresolved challenge and carefully reconsider the exponential material kinematics involved and the proper definition of the corresponding material functions. We show that by understanding the time-varying eigenvalues and eigenvectors of both the shearing kinematics and the evolving state of stress in the sample it is, in fact, possible to use an exponential shear test to quantitatively map out \textit{both} the full flow curve (of shear viscosity vs. shear rate) and the steady planar extensional viscosity curve as a function of deformation rate. These findings are supported by numerical simulations of a number of nonlinear viscoelastic constitutive models, asymptotic analysis, as well as experiments with several polymeric liquids.
\end{abstract}

\maketitle

\section{Introduction}

Extensional deformations are prevalent in many complex fluid processing operations \cite{petrie_extensional_2006,huang_when_2022}, therefore a rigorous characterization of a fluid's extensional rheology is crucial for material formulation and flow prediction. The tensile stress growth of a polymeric liquid can be particularly large in strong extensional flows, with stresses greatly exceeding those expected for a Newtonian fluid with the same shear viscosity \cite{metzner_stress_1970,baid_rheological_1977}. More generally, the extensional properties of a complex fluid are often very different from its shear properties and one cannot, in general, be predicted from the other \cite{petrie_elongational_1979}. Metzner \cite{metzner_stress_1970} emphasized that the ability of a constitutive equation to describe both shearing and extensional deformation is of ``singular importance in distinguishing between poor and adequate constitutive equations''. In contrast to shear measurements, however, which are routinely performed in a commercial rheometer, extensional measurements typically require specialized devices. There has been active research over the past decades on developing devices for measuring the extensional properties of complex fluids, ranging from counter-rotating fixtures for polymer melts \cite{sentmanat_miniature_2004} to filament stretching devices relying on fluid capillarity \cite{mckinley_filament-stretching_2002,dinic_extensional_2015,keshavarz_studying_2015}, to pressure-driven microfluidic devices \cite{haward_optimized_2012,haward_extensional_2023}. Each of these developments reflects a broader theme in Metzner's work: the need to characterize complex fluids under the actual strong flow histories relevant to their use, rather than characterizing their flow behavior only through linear viscoelastic or steady simple shear measurements.

Exponential shear was proposed by Doshi and Dealy \cite{doshi_exponential_1987} as a strong flow capable of efficiently stretching the microstructure of a complex fluid, which allows insight into the extensional rheological response of the material. The advantage is that this spatially homogeneous flow can be realized in a commercial rotational rheometer, which is often already being used for shear characterization of complex fluids and soft solids. In an ``ideal exponential flow'' \cite{doshi_exponential_1987}, the strain forcing function, parameterized by the values $A$ and $\alpha$, is $\gamma (t) = A \sinh(\alpha t)$, which increases exponentially with time. The corresponding strain rate is then readily calculated to be $\dot{\gamma} (t) = A \alpha \cosh(\alpha t)$. Z\"{u}lle \textit{et al.} \cite{zulle_deformation_1987} have also suggested using a ``nearly exponential flow'' in which $\gamma (t) = A (\exp(\alpha t) - 1)$, that some studies have used instead. In this case, the strain rate is $\dot{\gamma} (t) = A \alpha \exp(\alpha t)$. In these original papers, the strain and strain rate functions were multiplied by the adjustable constant pre-factor $A$, providing an additional degree of freedom. However, we show below that when the data is presented in terms of appropriate material functions, this pre-factor becomes inconsequential. In this paper, we focus on ``ideal exponential flow'', but later show that the difference between this and ``nearly exponential flow'' is not significant in the dynamical regime of interest, i.e. at long times when the growing exponential term dominates the function. Because material points separate apart exponentially in time under this protocol, similar to extensional flows, exponential shear has been suggested as an alternative characterization technique for investigating the extensional viscosity. Whether it truly mimics an extensional flow, however, has been hotly debated in the literature. 

Two material functions to analyze exponential shear were suggested by Doshi and Dealy \cite{doshi_exponential_1987} in their original paper, such that ``the behavior of different materials can be meaningfully compared and that deviations from linear viscoelastic behavior can be readily noted''. The first is the `exponential viscosity', corresponding to the instantaneous shear stress divided by the instantaneous strain rate. The authors plotted this material function against time, finding that, after an initial sharp overshoot, it decreases with time. Sivashinsky \textit{et al.} \cite{sivashinsky_new_1984} were the first to point out that plotting the exponential viscosity against the instantaneous strain rate approximates the steady shear viscosity flow curve. This was also noted by Venerus \cite{venerus_exponential_2000} and Neergaard \textit{et al.} \cite{neergaard_exponential_2000}, who further noted that the first normal stress coefficient can be approximated similarly. This was surprising since exponential shear is a transient protocol. However, these observations have not been thoroughly explained within a theoretical framework to establish their limits of validity. We analyze the kinematics of exponential shear flow in Section II, and demonstrate that this is because, at each instant of time during the exponential shearing deformation, the material attains a quasi-steady state. 

The second material function proposed by Doshi and Dealy \cite{doshi_exponential_1987} is the instantaneous principal stress difference $\Delta \sigma(t) = \sqrt{[N_1 (t)]^2 + 4 [\sigma_{12} (t)]^2}$ divided by twice the exponentially increasing strain rate $\dot{\gamma} (t) = A \alpha \cosh(\alpha t)$. This material function captures the ratio of the difference between the two eigenvalues of the stress tensor in shear, and the difference between the two eigenvalues of the strain rate tensor in shear. Doshi and Dealy proposed this as a material function to show strain hardening of the material, similar to observations under true extensional flows. However, it has been observed that this quantity is smaller than the stress growth coefficient measured in the start up of steady shear \cite{venerus_exponential_2000}, exhibiting strain softening by this definition and contrary to expectation.

Subsequent work has proposed various other material functions to elucidate the similarities between exponential shear and extensional flows, but there has been no agreement on a suitable material function that captures the strain hardening of a typical material subject to exponential shear. A few proposed approaches are listed here. Kwan, Woo and Shaqfeh \cite{kwan_experimental_2001} suggest dividing $\Delta \sigma$ by $\alpha$, assuming that the principal axes of stress are co-linear with the principal axes of the refractive index tensor. Moriam \textit{et al.} \cite{moriam_enhancing_2026} also applied this material function when using exponential shear to characterize cellulose-based textiles. Suneel \textit{et al.} \cite{suneel_characterisation_2003} suggest directly dividing the instantaneous shear stress by $\alpha$. Kroo \textit{et al.} \cite{kroo_unifying_2025} were the first to suggest correcting for the evolving deformation factor by scaling $\Delta \sigma$ using a rate factor based on the exact solution of the Upper Convected Maxwell model in exponential shear flow. 

However, it remains unclear whether exponential shear is closer to a simple shear or a planar extensional flow. While a number of authors believe that it can mimic an extensional flow \cite{doshi_exponential_1987,samurkas_strong_1989,kwan_brownian_1999,kwan_experimental_2001,kroo_unifying_2025}, others are more cautious and believe that it is not as effective at characterizing the extensional behavior of a material as true extensional flows \cite{graham_using_2001,suneel_characterisation_2003,wagner_exponential_2005}. Yosick \textit{et al.} \cite{yosick_kinetic_1997} have suggested that Large Amplitude Oscillatory Shear and exponential shear data can be used together to provide information on parameter identification for nonlinear constitutive models, further illustrating the utility of exponential shear. Yet other authors \cite{venerus_exponential_2000,neergaard_exponential_2000} have dismissed exponential shear as being ``no more capable of stretching polymer chains than is inception of steady shear at comparable instantaneous shear rates'' due to the vorticity present in the shearing kinematics. These discrepancies motivate a careful and rigorous re-examination of the kinematics of the flow, which is the focus of this paper. 

Notably, the majority of previous studies of exponential shear, such as those of Z\"{u}lle \textit{et al.} \cite{zulle_deformation_1987}, Samurkas \textit{et al.} \cite{samurkas_strong_1989} and Suneel \textit{et al.} \cite{suneel_characterisation_2003} do not measure the time-varying first normal stress difference $N_1 (t, \dot{\gamma}(t))$ and it is therefore not possible to evaluate the full time-evolving principal stress difference $\Delta \sigma (t)$ proposed by Doshi and Dealy \cite{doshi_exponential_1987}. We will later show that this is critically important in the correct interpretation of exponential shear rheometry. In fact, the only available experimental data sets containing both the time-evolving shear stress and the first normal stress difference under exponential shear are reported by Venerus \cite{venerus_exponential_2000} for a polymer melt and Kwan, Woo and Shaqfeh \cite{kwan_experimental_2001} for a Boger fluid, but both of these earlier reports only reach relatively low accumulated strains of no more than $\alpha t_\mathrm{max} \simeq 3$. Modern strain-controlled rheometers have very high torque and normal force sensitivities, and their high motor bandwidths enable access to much higher maximum strains (or angular displacements), prompting us to revisit the exponential shear technique here over a wide range of deformation rates. 

In this paper, our contributions are twofold. We first show, by analyzing the kinematics of the flow field, how exponential shear is fundamentally different from both simple shear and planar extension. This analysis hopefully resolves much of the confusion in the literature regarding where exponential shear is positioned with respect to two other canonical rheometric flows of steady shear and steady extension. Next, we demonstrate how dynamical measurements in exponential shear can be leveraged to access material functions that closely approximate those from \textit{both} simple shear and planar extension within a single rapid experiment. The key requirement is that \textit{both} the first normal stress difference and the shear stress have to be measured simultaneously. Under such conditions, we show that exponential shear is useful for obtaining rich rheological information on the nonlinear response of a complex fluid. 

The paper is organized as follows. We first examine the kinematics of exponential shear relative to simple shear and planar extension, and subsequently analyze two representative constitutive models for complex fluids. First we consider the quasi-linear Upper Convected Maxwell model, and then proceed to analyze the fully nonlinear Phan-Thien--Tanner model, showing in both cases that material functions representing both shearing and planar extensional flows can be obtained. Because single-mode constitutive models described by a single relaxation time are not completely representative of most real complex fluids (which are more typically characterized by a spectrum of relaxation times), we investigate how this additional complexity modifies the expected results. Finally, we use experimental data collected for two different complex fluids (a polydimethylsiloxane (PDMS) room-temperature entangled melt plus a semidilute polyisobutylene (PIB) solution in paraffin oil) to illustrate how exponential shear can be used to more fully characterize the extensional response of both polymeric melts and solutions, and potentially other classes of complex fluids. 

\section{Kinematics}

Exponential shear refers to a simple shearing flow with a strain rate that increases exponentially with time. Kroo \textit{et al.} \cite{kroo_unifying_2025} have shown that using the pre-factor $A = 2$ enables the imposed kinematics to closely mimic the time-dependent increase in the Hencky strain expected for a planar extensional flow. Throughout the present paper, the strain function $\gamma^e(t)$ and strain rate function $\dot{\gamma}^e (t)$ for an exponential shear flow protocol are hence taken to be:
\begin{subequations} \label{eq:expshear}
\begin{equation}
    \gamma^{~e} (t) = 2 \sinh (\alpha t)
\end{equation}
\begin{equation}
    \dot{\gamma}^{~e} (t) = 2 \alpha \cosh(\alpha t)
\end{equation}
\end{subequations}

Simple shear and planar extension differ in at least two ways. Firstly, material points separate apart algebraically in time for simple shear, but separate exponentially in time in planar extension. Secondly, the principal axes of the material stretch are fixed for planar extension, yet constantly evolve in time for simple shear. These two differences embody the competing arguments regarding the flow type classification of exponential shear. 

On the one hand, material points do indeed separate exponentially apart in time for an exponential shear flow. Various forms of this `strong flow' argument exist, and a well-known classification system proposed by Tanner and Huilgol \cite{tanner_classification_1975} suggests that a flow is considered strong if the strain history of a material element grows exponentially over time. This is quantified by the largest principal eigenvalue $\lambda_\mathrm{max}$ of the Finger strain tensor $\pmb{B}$, which must grow exponentially with time. In the field of nonlinear dynamics \cite{strogatz_nonlinear_2024}, the argument of the exponential function is also known as the largest Lyapunov exponent $\Lambda$, defined as:
\begin{equation}
    \Lambda = \lim_{t \rightarrow \infty} \left( \frac{\ln(\lambda_\mathrm{max}(\pmb{B}))}{t} \right)
\end{equation}
A strong flow as defined in this way implies that the Lyapunov exponent of the largest eigenvalue of the Finger strain tensor, which we refer to as the maximal Lyapunov exponent, must be finite. For planar extension at a fixed extension rate $\dot{\epsilon}_0$, the largest eigenvalue of the Finger strain tensor is $\exp(2\dot{\epsilon}_0 t)$, and the maximal Lyapunov exponent is thus $\Lambda = \lim_{t \rightarrow \infty} \left( \tfrac{\ln (\sqrt{\exp(2 \dot{\epsilon}_0 t}) )}{t} \right) = \dot{\epsilon}_0$, and is finite. For simple shear flow at a fixed shear rate $\dot{\gamma}_0$, and at a time $t$ corresponding to an accumulated strain $\gamma = \dot{\gamma}_0 t$, the largest eigenvalue of the Finger strain tensor is $\tfrac{\gamma + \sqrt{\gamma^2 +4}}{2}$, as first calculated by Lodge \cite{lodge_elastic_1964}. Substituting $\gamma (t) = \dot{\gamma}_0 t$, the maximal Lyapunov exponent in simple shear is thus $\Lambda = \lim_{t \rightarrow \infty} \left( \frac{\ln(\gamma + \sqrt{\gamma^2 +4})}{2t} \right) = 0$. For exponential shear, the largest eigenvalue of the Finger strain tensor is the same as for simple strain, $\tfrac{\gamma^e + \sqrt{[\gamma^e]^2 +4}}{2}$, but contains a time-varying strain history $\gamma^e (t) = 2 \sinh(\alpha t)$. The maximal Lyapunov exponent is then calculated to be $\Lambda = \alpha$, and is finite. This was also the kinematic argument set out by Doshi and Dealy \cite{doshi_exponential_1987} when they first proposed that exponential shear is a strong flow. 

On the other hand, Astarita \cite{astarita_objective_1979} has argued that the criterion of Tanner and Huilgol \cite{tanner_classification_1975} is limited only to motions with constant stretch history, and that a more general criterion is needed, based on the velocity gradient tensor. He hence proposed a materially objective criterion, which is now known as Astarita's D-criterion. The velocity gradient tensor $\nabla \pmb{v}$ is decomposed into two parts; a symmetric contribution given by the deformation rate tensor $\pmb{D} = \tfrac{1}{2} \pmb{\dot{\gamma}} = \tfrac{1}{2}(\nabla \pmb{v} + \nabla \pmb{v}^T)$ and an antisymmetric part, the spin tensor given by $\pmb{W} = \tfrac{1}{2} (\nabla \pmb{v} - \nabla \pmb{v}^T)$. Because the spin tensor $\pmb{W}$ is not frame indifferent, the rotation of the eigenvectors of the strain rate tensor, denoted by $\pmb{\Omega}$, must be subtracted from the spin tensor to obtain a materially objective tensor. Astarita \cite{astarita_objective_1979} proposed defining a relative spin tensor $\pmb{\bar{W}} = \pmb{W} - \pmb{\Omega}$, and calculating the following criterion:
\begin{equation}
    R_D = - \frac{\text{tr}(\pmb{\bar{W}}^2)}{\text{tr}(\pmb{D}^2)}
\end{equation}
This can be interpreted as the ratio of the amount of vorticity in the flow to the amount of stretching. For simple shear flow, $R_D = 1$, and for extensional flows, $R_D = 0$. However, for pure solid-body rotation, in which there is no stress due to stretching, Astarita's ratio gives $R_D \rightarrow \infty$, which is ill-posed and leads to numerical instabilities when used as a flow classification criterion in numerical simulations. Mompean \textit{et al.} \cite{mompean_general_2003} hence re-expressed this criterion in a normalized form:
\begin{equation}
    \xi = \frac{1 - R_D}{1 + R_D}
\end{equation}

By this definition, $\xi = 0$ for simple shear and $\xi = 1$ for extensional flows. However, $\xi = 0$ for exponential shear since it is also a (time-dependent) shear flow. Venerus \cite{venerus_exponential_2000} and Neergaard \textit{et al.} \cite{neergaard_exponential_2000} have used this to reason that the presence of vorticity in exponential shear necessarily means that it is a weak flow. Their argument is that the deforming material microstructure is able to rotate away from the principal axes of deformation to protect itself from further stretching, and hence is not stretched as efficiently as it otherwise would be in a true extensional flow. 

\begin{figure*}
\centering
\includegraphics[width=\textwidth]{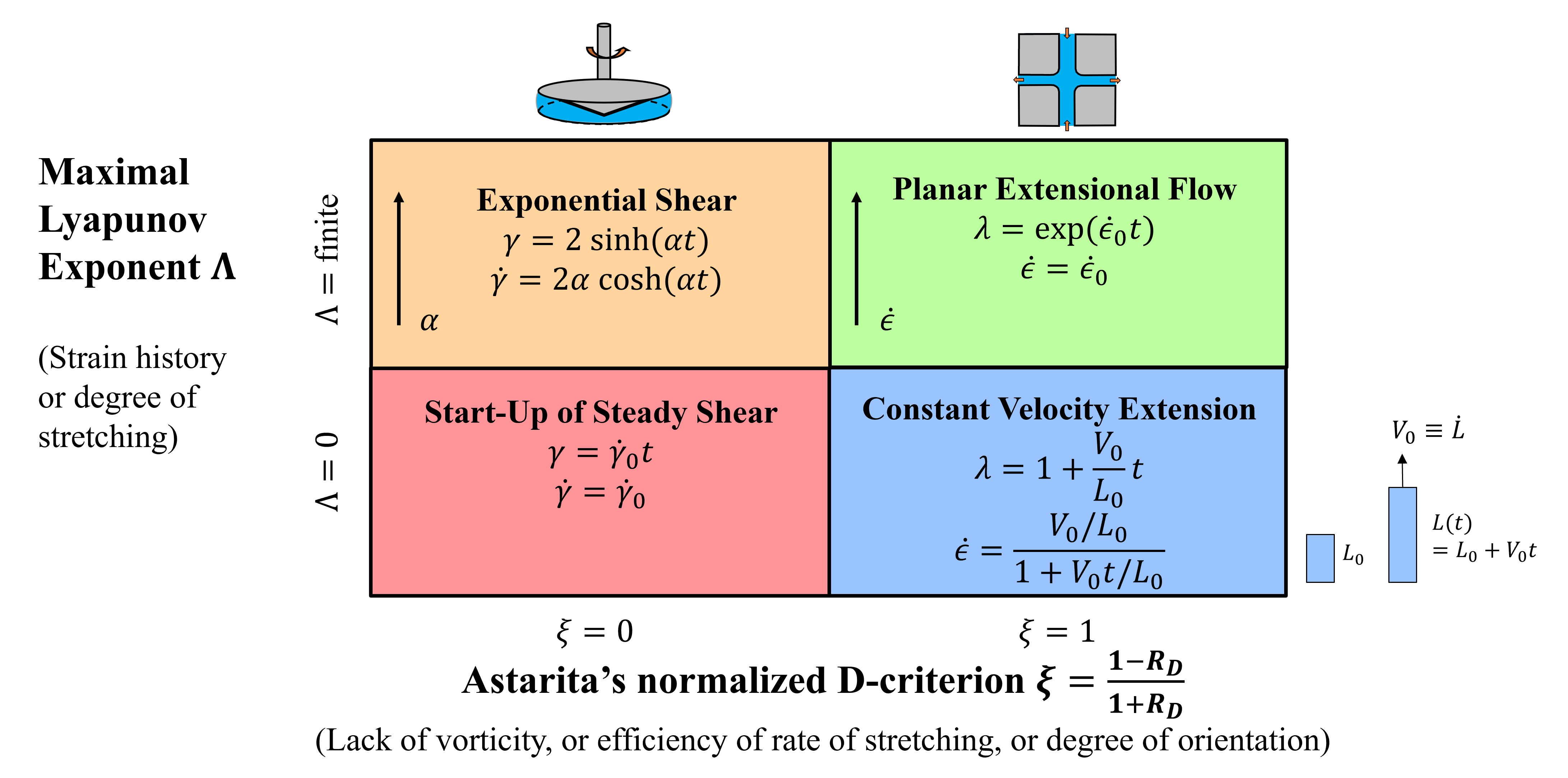}
\caption{A block diagram showing four types of flows or deformations, arranged according to the value of the maximal Lyapunov exponent and Astarita's normalized D-criterion. Steady shear and exponential shear both have strain $\gamma$ and strain rate $\dot{\gamma}$ while planar extensional flow and constant velocity extension are characterized by a stretch $\lambda$ and extension rate $\dot{\epsilon}$. For steady shear, the imposed strain rate is constant, $\dot{\gamma} = \dot{\gamma}_0$. For exponential shear, the imposed strain and strain rate vary as per Eq. \ref{eq:expshear} in the main text with a parameter $\alpha$, which has units [s$^{-1}$]. For planar extension, the imposed extension rate is constant, $\dot{\epsilon} = \dot{\epsilon}_0$. For constant velocity extension, the imposed stretch and extension rate vary with the constant velocity of stretching $V_0$ and the initial length of sample $L_0$.}
\label{fig:blockdiagram}
\end{figure*} 

From the above discussion, it is clear that exponential shear combines features of simple shearing and planar extensional flows, which partly explains why this flow protocol remains controversial and a source of confusion in the literature. More broadly, ideas about what constitutes a strong or weak flow, and how to express such criteria in a materially objective manner, continue to be debated even today \cite{poole_inelastic_2023}. Nevertheless, based on these two criteria as listed above, we can assert that exponential shear sits somewhere in-between the two limiting cases of simple shear and planar extension, and we express these three flows in a block diagram in Fig. \ref{fig:blockdiagram}. We will return to the fourth and last piece of the block diagram, which Petrie \cite{petrie_one_2006} has termed `Constant Velocity Extension', after an analysis of exponential shear.

\begin{figure*}
\centering
\includegraphics[width=0.9\textwidth]{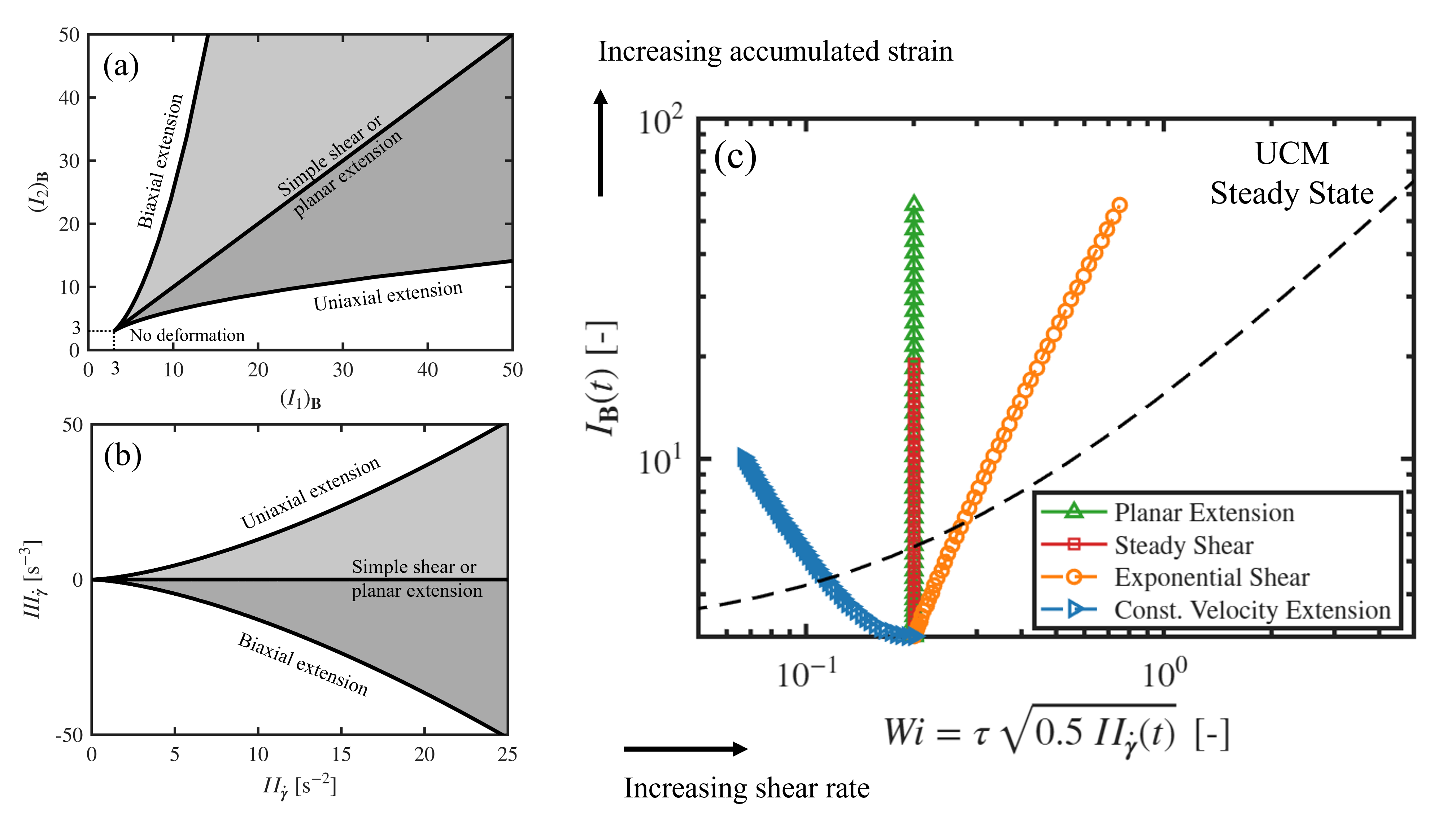}
\caption{(a) A graph showing the evolution of the first and second principal invariants of the Finger strain tensor first proposed by Sawyers \cite{sawyers_possible_1977}. (b) A graph of the second and third scalar invariants of the strain rate tensor \cite{macosko_rheology_1994}. (c) A kinematic cross-plot showing the evolution of the first principal invariant of the Finger strain tensor against the second scalar invariant of the strain rate tensor. The kinematic trajectories for the four classes of flows considered in Fig. \ref{fig:blockdiagram} are plotted from $t/\tau = 0$ to $t/\tau = 20$. For steady shear, $\dot{\gamma}_0 \tau = 0.2$; for planar extension, $\dot{\epsilon}_0 \tau = 0.1$; for exponential shear, $\alpha \tau = 0.1$; for constant velocity extension, $V_0 \tau /L_0 = 0.1$. A dashed line indicates the strain and strain rate at a time corresponding to five times the relaxation time after a step strain rate has been imposed for the UCM model. The region to the left of the line correspond to the locus of material points for a UCM fluid after a steady state flow has been attained.} 
\label{fig:invariantmap}
\end{figure*}

Without getting further drawn into the debate of what constitutes a strong or weak flow, we analyze the kinematics of exponential shear from a more fundamental point of view \cite{bird_dynamics_1987}. All possible flows and deformations can be expressed in terms of the strain rate tensor $\pmb{\dot{\gamma}} = 2 \pmb{D} = \nabla \pmb{v} + \nabla \pmb{v}^T$ and the Finger strain tensor $\pmb{B}$. Since these are rank-2 tensors, they are each uniquely defined by three scalar invariants. Taking the Finger strain tensor as an example, one such basis is the set of scalar invariants $\{I_{\pmb{B}} = \text{tr} (\pmb{B}), II_{\pmb{B}} = \text{tr} (\pmb{B}^2), III_{\pmb{B}} = \text{tr} (\pmb{B}^3) \}$. Another basis is the set of principal invariants $\{I_{1,\pmb{B}} = \text{tr} (\pmb{B}), I_{2,\pmb{B}} = \tfrac{1}{2} ((\text{tr} (\pmb{B}))^2 - \text{tr}(\pmb{B}^2)), I_{3,\pmb{B}} = \text{det} (\pmb{B}) \}$ which are related to each other through the Cayley-Hamilton Theorem. Assuming incompressibility of the material, $I_{\pmb{\dot{\gamma}}} = \text{tr}~\pmb{\dot{\gamma}} = 0$ and also $I_{3,\pmb{B}} = \text{det}~\pmb{B} = 1$. The remaining two invariants for each tensor are typically plotted in two graphs that show the limits of all physically realizable flows and deformations. The graph of $I_{2,\pmb{B}}$ against $I_{1,\pmb{B}}$ (Fig. \ref{fig:invariantmap}(a)) was first presented by Sawyers \cite{sawyers_possible_1977}, and he showed that all possible incompressible material deformations are bounded by uniaxial and biaxial extension. The case of no deformation corresponds to the point with coordinates (3,3) on the graph. This graph also shows that simple shear and planar extension are kinematically equivalent and lie on the same diagonal line. Similarly, the graph of $II_{\pmb{\dot{\gamma}}}$ against $III_{\pmb{\dot{\gamma}}}$ (Fig. \ref{fig:invariantmap}(b)) can be found in various forms \cite{macosko_rheology_1994,kroo_unifying_2025}, with one specific representation popularized by Chong, Perry and Cantwell \cite{chong_general_1990} for analyzing turbulent flows. In this case, all possible flows are again bounded by uniaxial and biaxial extension, and simple shear and planar extension again lie on the same line, which is horizontal in this map. 

Since nearly all flows performed in a rheometer (i.e. planar flows) are simple shear flows that are co-linear with planar extension in terms of their kinematics, the two graphs plotted separately are not particularly illuminating. 
% The only possible exceptions include orthogonal superposition or squeezing flows. 
We therefore compare and contrast the four flows summarized in Fig. \ref{fig:blockdiagram} by cross-plotting the two invariants $I_{1,\pmb{B}}$ and $II_{\pmb{\dot{\gamma}}}$, as shown in Fig. \ref{fig:invariantmap}(c). Note that because $I_{1,\pmb{B}}$ is dimensionless but $II_{\pmb{\dot{\gamma}}}$ has units of [s$^{-2}$], we define a dimensionless flow strength or Weissenberg number as $Wi = \tau \left(\tfrac{1}{2} II_{\pmb{\dot{\gamma}}}(t) \right)^{1/2}$, in which $\tau$ is a characteristic relaxation time of the test fluid under consideration. This now reveals the differences between the kinematics of the different flow protocols, without consideration of the rheological response of a specific material. The full expressions for the kinematics of each flow are given in Table \ref{tab:kinematics}. 

For a constant extension rate $\dot{\epsilon} = \dot{\epsilon}_0$ and shear rate $\dot{\gamma} = 2\dot{\epsilon}_0$ (selected to produce the same value of $II_{\pmb{\dot{\gamma}}}$), the start-up of steady shear falls on exactly the same vertical kinematic trajectory as planar extension, the only difference being that the trajectories for material points experiencing planar extension are spaced exponentially further apart while those for simple shear grow quadratically in time (because $I_{1,\pmb{B}} = 3 +\gamma^2 = 3+(\dot{\gamma}_0 t)^2 \sim t^2$). We analyze this response in more detail using the Upper Convected Maxwell (UCM) model, which is the universal limit of most complex fluids subject to low strain rate ($Wi \lesssim \mathcal{O}(1)$) forcing. The UCM model is given by the following equation:
\begin{equation} \label{ucmmodel}
    \tau \accentset{\nabla}{\pmb{\sigma}} + \pmb{\sigma} = \eta_0 \dot{\pmb{\gamma}}
\end{equation}
where the inverted triangle indicates the upper convected derivative $\accentset{\nabla}{\pmb{\sigma}} \equiv \frac{D \pmb{\sigma}}{D t} - (\nabla \pmb{v})^T \cdot \pmb{\sigma} - \pmb{\sigma} \cdot \nabla \pmb{v}$, and is parameterized by the viscosity $\eta_0$, and relaxation time $\tau$. For the start-up of steady shear, with a prescribed strain rate $\dot{\gamma} = \dot{\gamma}_0 \mathcal{H}(t)$, in which $\mathcal{H}(t)$ is the Heaviside step function, the stress responses are:
\begin{subequations} \label{ucmstartup}
\begin{equation}
    \sigma_{12} = \eta_0 \dot{\gamma}_0 \bigl(1 - \exp(-t/\tau) \bigr)
\end{equation}
\begin{equation}
    N_1 = 2 \eta_0 \tau \dot{\gamma}_0^2 \bigl(1 - (1 + t/\tau) \exp(-t/\tau) \bigr)
\end{equation}
\end{subequations}

If we assume that a time of five times the relaxation time (i.e. $t \simeq 5\tau$) is sufficient to attain steady state for a UCM fluid, we can plot a dashed black line in Fig. \ref{fig:invariantmap}(c) to demarcate when a steady state response might be expected. Any kinematic conditions to the left of the dashed line correspond to essentially a steady state response from the UCM fluid. Because the dynamical fluid response eventually reaches steady state at long times (and large accumulated strain), both steady shearing and planar extensional flows subject to the same flow strength (as quantified by the value of $II_{\pmb{\dot{\gamma}}}$ or the Weissenberg number of the flow) are expected to result in the same rheological response, which can be revealed by carefully considering the time-evolving principal axes of strain rate and stress. This formed the basis for the kinematic and rheological equivalence between steady shearing and planar extensional flows recently established by King and McKinley \cite{king_kinematic_2026}.

Plotting the kinematic trajectory of exponential shear on the same axes (orange open circles) reveals that it is \textit{not} the same as simple shear or planar extension. The trajectory of exponential shear on this log-log plot is instead a diagonal line with a positive slope of unity at exponentially increasing intervals in time, given by the expression $I_{1,\pmb{B}} = \tfrac{II_{\pmb{\dot{\gamma}}}}{2\alpha^2} - 1$. Changes in the value of $\alpha$ only shift the curve vertically on this log-log plot, but do not affect its slope. Hence, even though material points are indeed separating exponentially in time, the applied kinematics experienced by material elements are not exactly the same as either simple shear or planar extension. Importantly, because the trajectory of exponential shear is diagonal in this $I_{1,\pmb{B}}$--$II_{\pmb{\dot{\gamma}}}$ space, a material element experiences a progressively increasing strain rate as it is stretched. The diagonal trajectory also enters the quasi-steady state regime demarcated by the black dashed line. This indicates that once sufficiently high strains have been achieved (i.e. $I_{1,\pmb{B}} \gg 1$), we are able to obtain information on both \textit{steady} shearing and planar extensional flows across multiple applied strain rates \textit{within a single experiment}. We thus suggest that it is more appropriate to use exponential shear as a means to probe the evolving steady state behavior of a complex fluid subject to a \textit{range} of simple shearing or planar extensional flow states at different flow strengths, rather than interpreting the deformation history in terms of the transient material response at a \textit{single} shear or extension rate, since the kinematics of the flow with respect to time are markedly different from either of them. 

Previous authors \cite{kwan_experimental_2001,kroo_unifying_2025} have defined the Weissenberg number for exponential shear as $Wi = \alpha \tau$. However, as we show below, this product can also be viewed as a measure of the flow unsteadiness since $\alpha$ sets the timescale over which the deformation rate varies. Another preferred definition of a Weissenberg number is based on the time-varying shear rate $\dot{\gamma}(t)$ or extension rate $\dot{\epsilon}(t)$ i.e. $Wi = \dot{\gamma} \tau$ or $Wi = \dot{\epsilon} \tau$. We will use this latter definition of flow strength since we have established that exponential shear passes through a series of quasi-steady states and hence through a range of progressively increasing Weissenberg numbers. This is also consistent with the definition used above to plot the dimensionless abscissa of the kinematic trajectory map (Fig. \ref{fig:invariantmap}(c)). 

For exponential shear flow specifically, we can define a characteristic flow time scale for measuring the flow unsteadiness as $t_\mathrm{flow} \equiv \gamma^e(t) / \dot{\gamma}^e (t) = \tanh(\alpha t) / \alpha$, leading to a definition of the Deborah number in exponential shear flow as $De \equiv \tau / t_\mathrm{flow} = \alpha \tau \coth(\alpha t)$. At small times, exponential shear, like start-up of steady shear flow, is a weak but rapidly varying flow with $De \rightarrow \infty$, while $Wi(t=0) = 2\alpha \tau$. Conversely at long times $De(t \rightarrow \infty) =\alpha \tau $, while $Wi \sim \alpha\tau \exp(\alpha t)$. %\rightarrow \infty$. 
A Pipkin diagram \cite{ewoldt_mapping_2017} for any rheometric shear flow can be constructed with the Deborah number on the abscissa and the Weissenberg number on the ordinate. For a single exponential shear experiment, trajectories move from the lower right to the upper left of the Pipkin diagram. The dimensionless product $\alpha \tau$ provides a characteristic measure of \textit{both} the flow strength (at early times) and the flow unsteadiness (at long times). For clarity, henceforth we consistently refer to this product as $\alpha \tau \equiv De_{\infty}$ and use the following definition of the (time-dependent) Weissenberg number $Wi = \tau \dot{\gamma}(t)$.

\begin{table*}
\footnotesize
\caption{\label{tab:kinematics} Kinematics of the four classes of planar flows and deformations considered in the present work. Note that based on how the strain is defined, for simple shear flow, the strain rate $\dot{\gamma} = \tfrac{d \gamma}{dt}$, but for extensional flows, the extension rate $\dot{\epsilon} = \tfrac{d (\ln~\lambda)}{dt}$. The commonly used Hencky strain is $\epsilon = \ln(\lambda)$ and $\lambda(t)$ is the principal stretch.}
\begin{ruledtabular}
\begin{tabular}{cccccc}
Flow Type & $\pmb{B}$ & $\pmb{\dot{\gamma}}$ & $I_{\pmb{B}}=II_{\pmb{B}} = \text{tr}~\pmb{B}$ & $II_{\pmb{\dot{\gamma}}}=III_{\pmb{\dot{\gamma}}} = \text{tr}~\pmb{\dot{\gamma}^2}$ & Notes \\
\hline
Simple Shear & $\begin{pmatrix}
        1 + \gamma^2 & \gamma & 0 \\
        \gamma & 1 & 0 \\
        0 & 0 & 1
    \end{pmatrix}$ & $\begin{pmatrix}
        0 & \dot{\gamma} & 0 \\
        \dot{\gamma} & 0 & 0 \\
        0 & 0 & 0
    \end{pmatrix}$ & $3 + \gamma^2$ & $2\dot{\gamma}^2$ & \makecell{where $\gamma(t) = \dot{\gamma}_0 t$ \\ and $\dot{\gamma} = \dot{\gamma}_0$} \\
Planar Extension & $\begin{pmatrix}
        \lambda^2 & 0 & 0 \\
        0 & \lambda^{-2} & 0 \\
        0 & 0 & 1
    \end{pmatrix}$ & $\begin{pmatrix}
        2 \dot{\epsilon} & 0 & 0 \\
        0 & - 2 \dot{\epsilon} & 0 \\
        0 & 0 & 0
    \end{pmatrix}$ & $\lambda^2 + \lambda^{-2} + 1$ & $8\dot{\epsilon}^2$ & \makecell{where $\lambda(t) = \exp(\dot{\epsilon}_0 t)$ \\ and $\dot{\epsilon} = \dot{\epsilon}_0$}\\
Exponential Shear & $\begin{pmatrix}
        1 + \gamma^2 & \gamma & 0 \\
        \gamma & 1 & 0 \\
        0 & 0 & 1
    \end{pmatrix}$ & $\begin{pmatrix}
        0 & \dot{\gamma} & 0 \\
        \dot{\gamma} & 0 & 0 \\
        0 & 0 & 0
    \end{pmatrix}$ & $3 + \gamma^2$ & $2\dot{\gamma}^2$ & \makecell{where $\gamma(t) = 2 \sinh(\alpha t)$ \\ and $\dot{\gamma}(t) = 2\alpha \cosh(\alpha t)$} \\
\makecell{Constant Velocity \\ Extension} & $\begin{pmatrix}
        \lambda^2 & 0 & 0 \\
        0 & \lambda^{-2} & 0 \\
        0 & 0 & 1
    \end{pmatrix}$ & $\begin{pmatrix}
        2 \dot{\epsilon} & 0 & 0 \\
        0 & - 2 \dot{\epsilon} & 0 \\
        0 & 0 & 0
    \end{pmatrix}$ & $\lambda^2 + \lambda^{-2} + 1$ & $8\dot{\epsilon}^2$ & \makecell{where $\lambda(t) = 1 + \frac{V_0}{L_0} t$ \\ and $\dot{\epsilon}(t) = \frac{V_0/L_0}{1+V_0 t/L_0}$}\\
\end{tabular}
\end{ruledtabular}
\end{table*}

For completeness, we now address the last quadrant of the block diagram in Fig. \ref{fig:blockdiagram}, which Petrie \cite{petrie_one_2006} has termed `Constant Velocity Extension'. This is a type of flow in which, even though the principal axes of stretch are co-linear with the stress, material points only separate algebraically in time. Hence, similar to simple shear, the accumulated strain at long times $t \gg L_0 / V_0$ grows quadratically with time as well (because $I_{1,\pmb{B}} = [\lambda(t)]^2 = (1 + V_0t/L_0)^2 \sim t^2$). This mode of deformation is commonly performed for tensile testing of solid samples of initial length $L_0$, with the ends separating at a constant velocity $V_0$. The stretch grows as $\lambda(t) = 1 + \tfrac{V_0 t}{L_0}$, in which $V_0$ is the velocity of the separating end fixtures, and $L_0$ is the initial length of the sample. One of the few cases in which such a deformation history has been applied to a polymeric liquid is by Vinogradov \textit{et al.} \cite{vinogradov_viscoelastic_1970}  Petrie \cite{petrie_one_2006}, when discussing exponential shear in his review of extensional flows, notes that it is unclear whether this `constant velocity extension' has similarities to simple shear, much in the same way that exponential shear has similarities to planar extension. We can now answer this question definitively by examining the kinematics of the flow as shown in the blue trajectory in Fig. \ref{fig:invariantmap}(c). Even though the stretch is algebraically increasing with time, the material is subject to progressively decreasing strain rates with time (i.e. a decreasing flow strength), causing it to follow a leftward trajectory in Fig. 2(c). Unlike exponential shear, however, the material takes a progressively longer time to reach high accumulated strains. Connelly \textit{et al.} \cite{connelly_notes_1979}, in attempting to produce a true uniaxial extension, held a cylindrical sample of a polymer melt at constant length, with one fixed end and stretched the other end at a constant velocity using a rotating shaft. The resulting inhomogeneous deformation rate that they measured also decreased with time and has similarities to this type of constant velocity extension. A homogeneous realization of such a flow is expected to be difficult to generate in a rheometer since it requires spatial homogeneity and minimal end effects in the sample (where the sample is clamped) just as in planar extension. It will also be difficult to measure the dynamical response since the force generated by the fluid sample will disappear at long times into the noise floor due to the decreasing deformation rate and the decreasing cross-sectional area of the sample. Furthermore, the flow gives less rheological information on the material since the deformation moves towards the linear response regime with time (i.e. to progressively lower Weissenberg numbers, implying that the material exhibits progressively more `Newtonian' flow behavior). We thus regard this flow protocol as nothing more than a curiosity from the viewpoint of rheological characterization.

\section{Choice of Appropriate Material Functions}

Material functions are typically defined in terms of the system’s response (i.e. the dynamical output) divided by a measure of the imposed deformation history (input). At leading order, they should thus scale out the effect of flow strength; for example, the definition of a viscosity as the measured shear stress divided by the imposed shear rate results in a constant value for a Newtonian fluid. A proper material function should also reduce in the appropriate limit of low Weissenberg numbers ($Wi \ll 1$) to well-recognized results from linear viscoelasticity. 

Based on these considerations, there are three material functions that can be used to describe exponential shear flow. The first is the `exponential viscosity' $\eta^e (\alpha,t)$, which can replicate the steady shear viscosity $\eta(\dot{\gamma})$, as first proposed by Doshi and Dealy \cite{doshi_exponential_1987}. Similarly, the `exponential first normal stress coefficient' $\Psi_1^e (\alpha,t)$ can also be measured. The following material functions are to be plotted against the instantaneous strain rate $\dot{\gamma}^e(t)$, as defined by Eq. \ref{eq:expshear}(b):
\begin{subequations} \label{expviscosities}
\begin{equation} \label{expviscosity}
    \eta^e (\alpha, t) = \frac{\sigma_{12}(t)}{\dot{\gamma}^e(t)}
\end{equation}
\begin{equation} \label{exppsi1}
    \Psi_1^e (\alpha, t) = \frac{N_1 (t)}{[\dot{\gamma}^e(t)]^2}
\end{equation}
\end{subequations}

The last definition we introduce is a new composite material function $\eta_{[P1]}^e (\alpha, t)$ that combines independent measurements of $\sigma_{12} (t)$ and $N_1 (t)$ and can replicate the steady planar extensional viscosity $\eta_{P1}(\dot{\epsilon})$. This material function was recently proposed by King and McKinley \cite{king_kinematic_2026}, and is to be plotted against the effective extension rate $\dot{\epsilon}_\mathrm{eff}^e(t)$:
\begin{subequations} \label{compositematerialfunction}
\begin{equation}
    \eta_{[P1]}^e (\alpha, t) = \frac{\Delta \sigma (t)}{\dot{\epsilon}_\mathrm{eff}^e (t)} = 4 \eta^e(t) \left(1 + \left(\frac{N_1(t)}{2\sigma_{12}(t)} \right)^2 \right)
\end{equation}
\begin{equation}
    \dot{\epsilon}_\mathrm{eff}^e (\alpha, t) = \frac{\sigma_{12}(t) \dot{\gamma}^e(t)}{\Delta \sigma(t)} = \frac{\sigma_{12}(t) \dot{\gamma}^e(t)}{\sqrt{[N_1(t)]^2 + 4[\sigma_{12}(t)]^2}} = \frac{\dot{\gamma}^e(t)}{2 \sqrt{1 + \left(\frac{N_1(t)}{2 \sigma_{12}(t)}\right)^2}}
\end{equation}
\end{subequations}

We provide a short overview of this new composite material function here. Because of the presence of vorticity in any shearing flow (including both steady shear and exponential shear), the principal axes of the stress tensor are constantly evolving with time. However, in a shear flow, the axes of the strain rate tensor are fixed at $\pm 45$ degrees to the flow direction. The magnitude of the eigenvalues of the strain rate tensor are constant in a steady shear flow, but increase (exponentially) in a strong flow such as exponential shear. By contrast, in an extensional flow, the strain rate tensor is fixed in time and coincident with the principal axes of the stress tensor, and it is only the magnitudes of the principal stresses that evolve with time as the material microstructure is deformed.

In order to remove the rotation from the kinematics, the key insight reported in our previous paper \cite{king_kinematic_2026} was to project the strain rate tensor onto the principal axes of the time-evolving stress tensor. This then produces an ``effective extension rate'' for exponential shear that can be used to define a material function which can be fairly compared with the response measured in planar extensional flow. We note here that in planar extension, there are two planar extensional viscosities due to the asymmetry induced in the Cartesian system. The second planar viscosity is small and seldom measured, but can in principle, also be measured if the second normal stress difference measured in the rheometer can be determined as well. We also note that to effectively use this formula, the solvent contribution to the total stress should be removed. Otherwise, the solvent contribution can dominate at high strain rates (due to the linearity between the Newtonian stress tensor and the flow strength) and the calculated extensional viscosity will be inaccurate. 

We note that the ratio $\tfrac{N_1(t)}{2\sigma_{12}(t)}$ appearing in Eq. \ref{compositematerialfunction} is a measure of the recoverable elastic shear, as defined by White \cite{white_dynamics_1964} in the original definition of the Weissenberg number. At short times (i.e. small strain rates), $N_1$ is negligibly small compared to the shear stress. The effective extension rate $\dot{\epsilon}_\mathrm{eff}^e$ defined in Eq. \ref{compositematerialfunction}(b) tends to a value of $\dot{\gamma}^e(t) / 2$. The composite material function $\eta_{[P1]}^e$ defined in Eq. \ref{compositematerialfunction}(a) approaches $4 \eta^e$, which is the expected Newtonian limit. At long times (i.e. larger strain rates), $N_1$ grows rapidly to become larger than the shear stress, and the ratio $\tfrac{N_1(t)}{2\sigma_{12}(t)}$ controls how both the effective extension rate $\dot{\epsilon}_\mathrm{eff}^e$ and the composite material function $\eta_{[P1]}^e$ evolve nonlinearly with time.

The next sections will explain how, for carefully chosen values of the exponential flow strength parameter $\alpha$, these material functions (Eq. \ref{expviscosities} and \ref{compositematerialfunction}) can be used to obtain the material functions that capture steady shear and planar extensional flows.

\section{Examples from Theoretical Models}

We illustrate the usefulness of exponential shear in probing the nonlinear properties of a complex fluid by first considering the quasilinear Upper Convected Maxwell model, and then a fully nonlinear model, the Phan-Thien--Tanner model. 

\subsection{Upper Convected Maxwell model}

The Upper Convected Maxwell (UCM) model is typically expressed as a differential tensorial equation and is one of the simplest possible materially objective constitutive equations for a polymeric liquid \cite{bird_dynamics_1987}. It is also equivalent to the Lodge rubbery liquid model, which is an integral equation derived initially through a transient network model for an elastic liquid. The UCM model was given in Eq. \ref{ucmmodel}, and has been shown in multiple papers \cite{doshi_exponential_1987,samurkas_strong_1989,kwan_experimental_2001,kroo_unifying_2025} to have the following analytical results in exponential shear:
\begin{subequations} \label{ucmexpshear}
\begin{equation}
    \sigma_{12} (t) = \frac{\eta_0 \alpha}{1 - \alpha^2 \tau^2} \left(\frac{\dot{\gamma}^e}{\alpha} - \alpha \tau \gamma^e - 2 e^{-t/\tau} \right)
\end{equation}
\begin{equation}
    N_1(t) = \frac{2\alpha^2 \tau \eta_0}{1 - \alpha^2 \tau^2} \left( \frac{1+2\alpha^2 \tau^2}{1 - 4\alpha^2 \tau^2} \left( \frac{[\gamma^e]^2}{2} + \frac{[\dot{\gamma}^e]^2}{2\alpha^2}\right) - \frac{3\tau \gamma^e \dot{\gamma}^e}{1 - 4 \alpha^2 \tau^2} - \frac{2\gamma^e e^{-t/\tau}}{\alpha \tau} - \frac{3 e^{-t/\tau}}{1 - 4 \alpha^2 \tau^2} - e^{-t/\tau} + 2 \right)
\end{equation}
\end{subequations}
where, as defined previously, $De_\infty = \alpha \tau$ is the long-time limit of the Deborah number and also a measure of the initial flow strength. It is clear that the UCM model in exponential shear exhibits rich dynamical behavior, and the material stress response is complicated by the fact that it never reaches a steady state or limit cycle, unlike typical rheological flow protocols. Nevertheless, it can be analyzed in various asymptotic limits. Demarquette and Dealy \cite{demarquette_nonlinear_1992} showed that a Taylor expansion in the limit that both $\alpha \tau \ll 1$ and $\alpha t \ll 1$ for the exponential viscosity $\eta^e(\alpha, t)$ defined in Eq. \ref{expviscosity} leads to the shear stress growth coefficient $\eta^+(t) = \eta_0(1 - \exp(-t/\tau))$. Sivashinsky \textit{et al.} \cite{sivashinsky_new_1984} found that when $\alpha$ was increased, there is a `bump' in the curve of the shear stress against time. At $t=0$, there is an initial finite step strain rate $\dot{\gamma}^e_0 = 2\alpha$ that is applied to the material. For small values of $\alpha$, the initial shear stress response can be predicted by the theory of linear viscoelasticity, and the first normal stress difference can also be obtained by retaining second order terms only, giving predictions which are exactly the same as Eq. \ref{ucmstartup}: 
\begin{equation}
    \lim_{\alpha t \ll 1} \sigma_{12}(t) = \eta^+ \dot{\gamma}^e_0 = \eta_0 \dot{\gamma}^e_0 \bigl(1 - \exp(-t/\tau) \bigr)
\end{equation}
\begin{equation}
    \lim_{\alpha t \ll 1} N_1 (t) = \Psi_1^+ [\dot{\gamma}^e_0]^2 = 2 \eta_0 \tau [\dot{\gamma}^e_0]^2 \bigl(1 - (1 + t/\tau) \exp(-t/\tau) \bigr)
\end{equation}

\begin{figure*}
\centering
\includegraphics[width=0.9\textwidth]{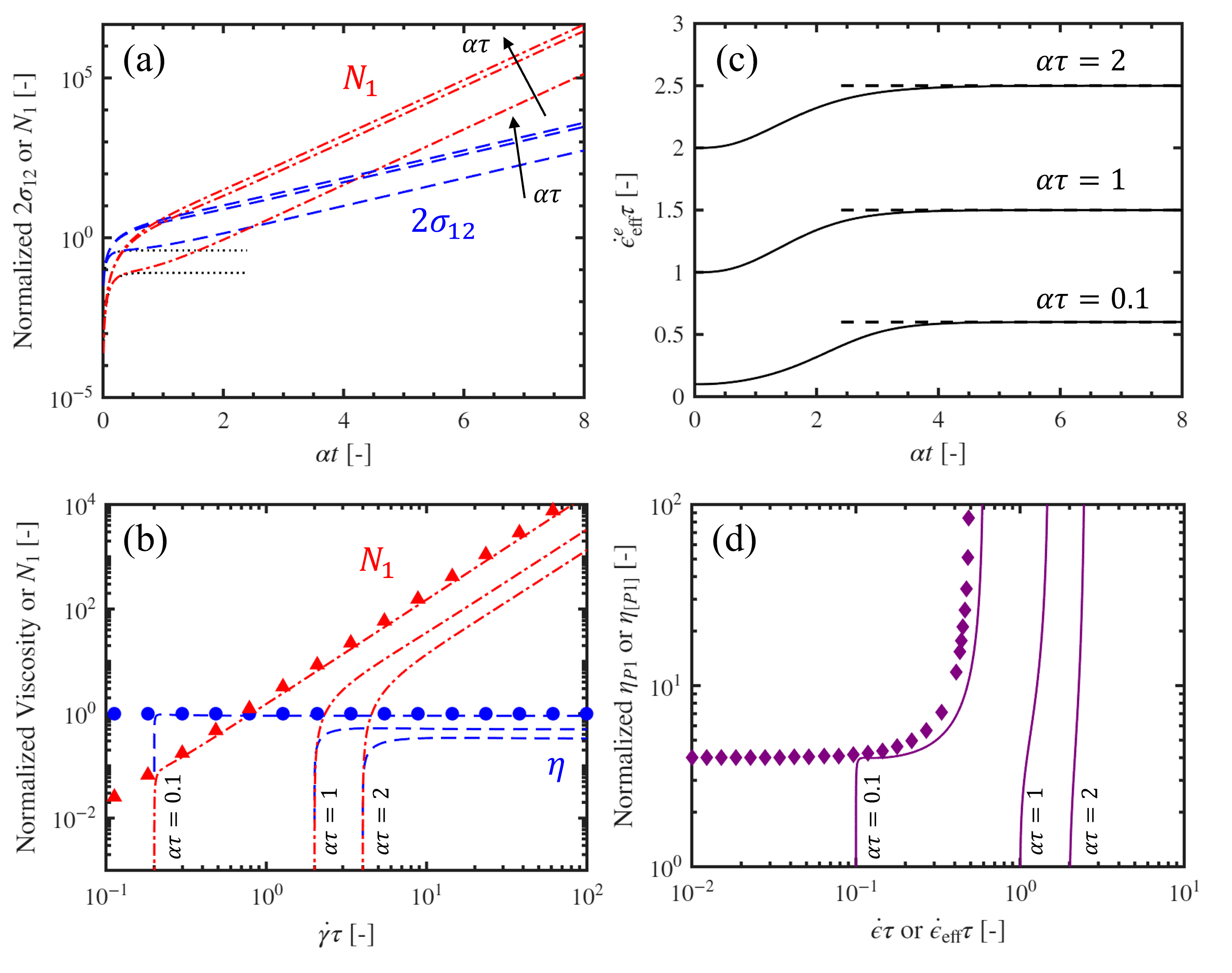}
\caption{Exponential shear flow of the Upper Convected Maxwell (UCM) model for different dimensionless flow strengths ($\alpha \tau = 0.1, 1, 2$). All stress quantities are normalized by a modulus $G = \eta_0 / \tau$, viscosities are normalized by the shear viscosity $\eta_0$, and shear and extension rates by $1/\tau$. All points are steady state data and lines are drawn for exponential shear. (a) The evolution of the shear stress and first normal stress difference against dimensionless time. Curves of increasing $\alpha \tau$ are from bottom to top. The leading order responses to an initial step strain rate for $\alpha \tau = 0.1$ are indicated by black dotted lines for both the shear stress and also for $N_1$. (b) Evaluation of the dimensionless steady shear viscosity and the first normal stress difference against dimensionless strain rate $Wi(t) = \tau \dot{\gamma}(t)$. Curves of increasing $\alpha \tau$ are arranged from left to right. (c) Evolution of the effective extension rate against dimensionless time. (d) The composite planar extensional viscosity function against effective extension rate, which diverges at different critical extension rates depending on the value of $\alpha \tau$. Curves of increasing $\alpha \tau$ are arranged from left to right.} 
\label{fig:ucm}
\end{figure*}

This response is shown in Fig. \ref{fig:ucm}(a) where the initial step strain rate response is indicated by a dotted line. Since linear viscoelasticity is a first order dynamical process, transient effects tend to last about 5 multiples of the relaxation time. Thereafter, the rapid increase in the strain rate leads to other nonlinear mechanisms taking over. A large value of $\alpha$ leads to both a large initial step strain rate, and also a very rapid increase in the strain rate, which leads to the convolution of both the linear viscoelastic response and more nonlinear processes that appear at higher strain rates. 

At longer times $\alpha t \gg 1$, Demarquette and Dealy \cite{demarquette_nonlinear_1992} note that the ``exponential viscosity is always approaching but lagging behind the viscosity, which is decreasing rapidly due to the exponential increase in shear rate''. This is because the linear viscoelastic response is eclipsed by the nonlinear response. We isolate the fastest growing terms in Eq. \ref{ucmexpshear} for $\alpha t \gg 1$, and find that:
\begin{equation} \label{ucmexpshear_shearstress}
    \lim_{\alpha t \gg 1} \sigma_{12}(\alpha, t) = \frac{\eta_0 \alpha}{1 + \alpha \tau} \exp(\alpha t)
\end{equation}
\begin{equation} \label{ucmexpshear_n1}
    \lim_{\alpha t \gg 1} N_1 (\alpha, t) =  \frac{(2\eta_0 \alpha) (\alpha \tau)}{(1+ \alpha\tau)(1 + 2\alpha\tau)} \exp(2\alpha t)
\end{equation}

It is clear that the elastic normal stress grows faster in time than the shear stress. As we noted in Section III, the ratio $\tfrac{N_1 (t)}{2 \sigma_{12} (t)}$ plays an important role in understanding the dynamics of exponential shear. Kroo \textit{et al.} \cite{kroo_unifying_2025} have denoted the dimensionless time at which the curves of $2\sigma_{12} (t)$ and $N_1 (t)$ cross over in Fig. \ref{fig:ucm}(a) as $\alpha t^*$. For the UCM model, they determined asymptotic limits for the value of $\alpha t^*$ depending on the value of $\alpha \tau$ applied, indicating that $\alpha t^*$ is a good measure of the elasticity within the material. In the limit of $\alpha \tau \gg 1$, they showed for the UCM model that $\alpha t^* = \sinh^{-1}(1) \approx 0.881$.

These stress functions Eq. \ref{ucmexpshear_shearstress} and \ref{ucmexpshear_n1} can then be used to evaluate the material functions (Eq. \ref{expviscosity} and \ref{exppsi1}), which turn out to be time-independent:
\begin{equation} \label{ucmexpshear_viscosity}
    \lim_{\alpha t \gg 1} \eta^e (\alpha) = \lim_{\alpha t \gg 1} \frac{\sigma_{12} (\alpha)}{\dot{\gamma}^e(\alpha)} = \frac{1}{1 + \alpha \tau} \eta_0
\end{equation}
\begin{equation} \label{ucmexpshear_psi1}
    \lim_{\alpha t \gg 1} \Psi_1^e (\alpha) = \lim_{\alpha t \gg 1} \frac{N_1(\alpha)}{[\dot{\gamma}^e(\alpha)]^2} = \frac{1}{(1+ \alpha\tau)(1 + 2\alpha\tau)} (2\eta_0 \tau) = \frac{1}{(1+ \alpha\tau)(1 + 2\alpha\tau)} \Psi_{1,0}
\end{equation}

Since the exponential viscosity $\eta^e$ for the UCM model differs from the steady shear viscosity $\eta_0$ by a factor of $(1 + \alpha \tau)^{-1}$,  $\eta^e$ decreases for higher values of $\alpha \tau$, therefore explaining the observation of Demarquette and Dealy \cite{demarquette_nonlinear_1992}. This is also shown in Fig. \ref{fig:ucm}(b), in which better agreement between the two material functions is observed for smaller values of $\alpha$. To obtain exactly the same material functions as those given by steady shear, \textit{both} conditions $\alpha t \gg 1$ and $\alpha \tau \ll 1$ in exponential shear must be satisfied.

Next, we demonstrate the utility of our second proposed material function (Eq. \ref{compositematerialfunction}). Substitution of these expressions (Eq. \ref{ucmexpshear_viscosity} and \ref{ucmexpshear_psi1}) at long times into Eq. \ref{compositematerialfunction}(a) allows us to obtain an approximation to the planar extensional viscosity for the UCM model:
\begin{equation} \label{ucmexpshear_compositeviscosity}
    \lim_{\alpha t \gg 1} \eta_{[P1]}^e (\alpha) \simeq \frac{4 \eta_0}{(1 + \alpha \tau)} \frac{1}{\left(1 + \frac{2\dot{\epsilon}_\mathrm{eff}^e \tau}{1 + 2\alpha \tau}\right) \left(1 - \frac{2\dot{\epsilon}_\mathrm{eff}^e \tau}{1 + 2\alpha \tau}\right)}
\end{equation}

At long times, the effective extension rate defined in Eq. \ref{compositematerialfunction}(b) also saturates to a constant value:
\begin{equation}
    \lim_{\alpha t \gg 1} \dot{\epsilon}_\mathrm{eff}^e (t) = \frac{1 + 2\alpha \tau}{2\tau}
\end{equation}

In the limit of $\alpha \tau \ll 1$, we recover from Eq. \ref{ucmexpshear_compositeviscosity} the steady planar extensional viscosity as a function of the effective extension rate as shown in Fig. \ref{fig:ucm}(d). This gives the well-known Trouton ratio of $Tr = \eta_{P1} / \eta = 4$ at low extension rates, and which diverges at an extension rate of $1/(2\tau)$, corresponding to a coil-stretch transition \cite{de_gennes_coil-stretch_1974}. When plotted in dimensionless form as shown in Fig. \ref{fig:ucm}(c), it is clear that the dimensionless extension rate experienced by material elements evolves with time, beginning at a value $\tau \dot{\epsilon}_\mathrm{eff} (t=0) = \alpha \tau$ and saturating at a value of $\tfrac{1}{2} + \alpha \tau$.

From the analysis above, there are two criteria that must be met in an exponential shear flow to replicate steady shear and planar extension data. Firstly, we need to access long times $\alpha t \gg 1$, such that linear viscoelastic effects arising from the initial imposition of a step strain rate do not affect the measured material functions. Also, we seek to impose $\alpha \tau \ll 1$, for two reasons. Firstly, as revealed from our analysis of the material functions above, this minimizes the deviations from the true steady shear and planar extensional flow material functions. Secondly, a low value of $\alpha$ leads to a low initial strain rate (cf. Eq. \ref{eq:expshear}(b)), which enables us to recreate a larger portion of the flow curve. The downside of choosing a small value of $\alpha$ is that the experiments take longer to perform. For instance, to reach a final strain rate of $\dot{\gamma} = 1000$ s$^{-1}$, an experiment with $\alpha = 0.1$ s$^{-1}$ takes about 92 s to complete, but an experiment with $\alpha = 0.01$ s$^{-1}$ takes about 1150 s (or 19 mins) to complete, which may be similar in duration to a conventional flow sweep, depending on the equilibration time and number of points specified per decade. Nevertheless, the continuous recording of the shear stress and first normal stress difference during the exponential shear protocol allows access to an even greater number of independent data points compared to flow sweeps at discrete strain rates.

\subsection{Caveats for Real Multi-Mode Materials}

Two important points for real materials stand in contrast to our analysis of the UCM model. Firstly, Samurkas \textit{et al.} \cite{samurkas_strong_1989} found experimentally that the exponential viscosity $\eta^e$ ``practically `catches up' with the steady-state viscosity, i.e., at long times (high instantaneous shear rates)'' for a branched low-density polyethylene (LDPE) melt, as pointed out by Demarquette and Dealy \cite{demarquette_nonlinear_1992}. Venerus \cite{venerus_exponential_2000} has also reported this phenomenon for a different LDPE melt. This stands in contrast to the predictions of the UCM model (Eq. \ref{ucmexpshear_viscosity}), in which $\eta^e$ is always lower than the shear viscosity $\eta$. This observed difference for real fluids cannot be explained by the UCM model, but will be rationalized in the next section using the linear PTT model. 

Secondly, the linear viscoelastic response of most real fluids is characterized by a spectrum of relaxation times, rather than by a single dominant relaxation time $\tau$. Therefore, it is not straightforward to determine whether the criterion $\alpha \tau \ll 1$ is satisfied, since a discrete relaxation spectrum assigns a different value of $\tau$ to each relaxation mode. For the case of a UCM model with $n$ modes, the model predicts in steady shear flow that $\sigma_{12} (\dot{\gamma}) = \sum_{i=1}^n \eta_i \dot{\gamma}$ and $N_1 (\dot{\gamma}) = 2 \sum_{i=1}^n \eta_i \tau_i \dot{\gamma}^2$, in which $\eta_i$ and $\tau_i$ are the viscosity and relaxation time contributions of each $i^{\text{th}}$ mode. 

The steady shear viscosity is therefore:
\begin{equation}
    \eta \equiv \frac{\sigma_{12}}{\dot\gamma} = \sum_{i=1}^n \eta_i
\end{equation}

But the steady state limit of the exponential shear viscosity is:
\begin{equation}
    \lim_{\alpha t \gg 1} \eta^e = \sum_{i=1}^n \frac{1}{1 + \alpha \tau_i} \eta_i
\end{equation}

Similarly, we may consider the second order material response of the UCM model. The zero-shear rate limit of the steady first normal stress coefficient is:
\begin{equation}
    \Psi_1 = \sum_{i=1}^n 2 \eta_i\tau_i
\end{equation}

But the steady state limit of the exponential first normal stress coefficient is:
\begin{equation}
    \lim_{\alpha t \gg 1} \Psi_1^e = \sum_{i=1}^n \frac{1}{(1 + \alpha \tau_i)(1 + 2\alpha \tau_i)} 2 \eta_i \tau_i
\end{equation}

Just as we determined from our analysis of the single mode model, there is an offset between the simple shear and exponential shear predictions, but the magnitude of this offset will depend on the distribution of relaxation times. The true planar extensional viscosity $\eta_{P1}$ that we ultimately wish to determine is given by the multi-mode UCM model as:
\begin{equation}
    \eta_{P1} = \sum_{i=1}^n \frac{4\eta_i}{(1 + 2 \tau_i \dot{\epsilon})(1 - 2 \tau_i \dot{\epsilon})}
\end{equation}
The coil-stretch transition occurs at a critical Weissenberg number that corresponds to the \textit{longest} relaxation of the fluid, $Wi_{CS} = \dot{\epsilon}_{CS} \tau_1 = \tfrac{1}{2}$. However, the composite material function constructed from the total (measurable) shear stress and first normal stress difference shows a dependence on the viscosity-weighted relaxation time defined as:
\begin{equation} \label{viscosityweightedrelaxationtime}
    \bar{\tau}_\eta = \frac{\sum_{i=1}^n \eta_i \tau_i}{\sum_{i=1}^n \eta_i}
\end{equation}
Expressing $\dot{\gamma}$ in terms of $\dot{\epsilon}_\textrm{eff}$ from Eq. \ref{compositematerialfunction}, we obtain the relation $\dot{\gamma}^2 = 4 \dot{\epsilon}_\mathrm{eff}^2 / [(1 + 2\dot{\epsilon}_\mathrm{eff} \bar{\tau}_\eta)(1 - 2 \dot{\epsilon}_\mathrm{eff} \bar{\tau}_\eta)]$. The composite material function computed from steady shear measurements hence predicts a planar extensional viscosity:
\begin{equation}
    \eta_{[P1]} =  \frac{4 \sum_{i=1}^n\eta_i}{(1 + 2 \bar{\tau}_\eta \dot{\epsilon}_\mathrm{eff})(1 - 2 \bar{\tau}_\eta \dot{\epsilon}_\mathrm{eff})}
\end{equation}
The coil-stretch transition predicted by this composite material function thus occurs at a critical Weissenberg number that corresponds to the \textit{viscosity-weighted} relaxation time of the fluid, $Wi_{CS} = \dot{\epsilon}_{CS} \bar{\tau}_\eta = \tfrac{1}{2}$.

For measurements made in exponential shear characterized by an exponential rate $\alpha$, the composite material function we have proposed shows a dependence on a more complex weighted relaxation time defined as:
\begin{equation}
    \bar{\tau}_{\alpha} = \frac{\sum_{i=1}^n \frac{1}{(1+\alpha \tau_i)(1 + 2\alpha \tau_i)}\eta_i \tau_i}{\sum_{i=1}^n \frac{1}{1 + \alpha \tau_i} \eta_i}
\end{equation}
We can similarly express $\dot{\gamma}^e$ in terms of $\dot{\epsilon}_\textrm{eff}$ from Eq. \ref{compositematerialfunction}(b), to obtain the relation $[\dot{\gamma}^e]^2 = 4 \dot{\epsilon}_\mathrm{eff}^2 / [(1 + 2\dot{\epsilon}_\mathrm{eff} \bar{\tau}_\alpha)(1 - 2 \dot{\epsilon}_\mathrm{eff} \bar{\tau}_\alpha)]$. The composite material function using exponential shear hence predicts a planar extensional viscosity:
\begin{equation}
    \lim_{\alpha t \gg 1} \eta_{[P1]}^e =  \frac{4 \sum_{i=1}^n \frac{1}{1+\alpha \tau_i}\eta_i}{(1 + 2 \bar{\tau}_\alpha \dot{\epsilon}_\mathrm{eff})(1 - 2 \bar{\tau}_\alpha \dot{\epsilon}_\mathrm{eff})}
\end{equation}
The coil-stretch transition predicted now occurs at a critical Weissenberg number that corresponds to a more complex moment of the discrete relaxation time spectrum of the fluid that is dependent on $\alpha$, the viscosity $\eta_i$ and relaxation time $\tau_i$ for each relaxation mode, $Wi_{CS} = \dot{\epsilon}_{CS} \bar{\tau}_\alpha = \tfrac{1}{2}$.

Graham \textit{et al.} \cite{graham_using_2001} have attempted to fit the linear viscoelastic behavior of a Pom-Pom model using a multi-mode UCM model and using exponential shear data. They showed that this complicates fitting because some modes do not get stretched as much as others. The analysis above also shows that this is the case. In fact, when we apply the kinematic and rheological equivalence between steady shearing and planar extensional flows to the multi-mode UCM model, we get a coil-stretch transition not at the longest relaxation time, but rather at the viscosity-weighted relaxation time. Nevertheless, because real materials exhibit a spectrum of relaxation times, we would not expect a sharp coil-stretch transition at a specific extension rate, but rather, a broadening of the transition. Given the dearth of experimental data for planar extensional flows of a real fluid, we might be able to use exponential shear as a guide provided we fulfill the criterion $\alpha \bar{\tau}_\eta \ll 1$. In the final section on experimental data, we show that multiple relaxation modes may not be as restrictive a limitation as we might first assume.

\subsection{Linear Phan-Thien--Tanner Model}

We next show how this protocol works in the context of a fully nonlinear model. We choose a common broadly applicable phenomenological nonlinear model based on polymer network formation and destruction, the Phan-Thien--Tanner (PTT) model \cite{phan-thien_new_1977}. Specifically, we choose the linear variant of the PTT model with the slip parameter set to zero. The constitutive response is thus given by:
\begin{equation}
    \tau \accentset{\nabla}{\pmb{\sigma}} + \left(1 + \frac{\varepsilon \tau}{\eta_0}\text{tr}(\pmb{\sigma}) \right) \pmb{\sigma} = \eta_0 \dot{\pmb{\gamma}}
\end{equation}
where the model nonlinearity is controlled by a dimensionless parameter $\varepsilon$, which is a measure of the enhanced rate of rupture events in the transient entangled network, and the model simplifies to the quasilinear Upper Convected Maxwell (UCM) model (Eq. \ref{ucmmodel}) when $\varepsilon = 0$. 

Shogin \cite{shogin_full_2021} has calculated analytical expressions for the steady shear viscosity predicted by the linear PTT model. Because of their mathematical complexity, we will only use the asymptotic expansions at high shear rates for comparison. Asymptotic analysis of the linear PTT model under steady shear at large strain rates (refer to Eq. 38 and 39 in the paper by Shogin \cite{shogin_full_2021}) gives:
\begin{subequations} \label{shogin}
\begin{equation}
    \sigma_{12} \simeq \eta_0 (2 \varepsilon \tau^2)^{-1/3} \dot{\gamma}^{1/3} + \mathcal{O}(\dot{\gamma}^{-1/3})
\end{equation}
\begin{equation}
    N_1 \simeq \eta_0 (\varepsilon^{2} \tau / 2)^{-1/3} \dot{\gamma}^{2/3}  + \mathcal{O}(1)
\end{equation}
\end{subequations}

To explain why the exponential viscosity $\eta^e$ is able to replicate the steady shear viscosity $\eta(\dot{\gamma})$ measured in real fluids to a higher degree of accuracy than  the UCM model (as observed by Samurkas \textit{et al.} \cite{samurkas_strong_1989}) we use the ``method of dominant balances'' \cite{bender_advanced_1984}. This approximates the solution to the system of differential equations by solving a simplified form of the system that contains only the terms that dominate the solution whilst treating the other terms as minor corrections. The differential equations for the evolution of the stress components in the linear PTT model under simple shear are:
\begin{subequations}
\begin{align}
\begin{split}
    \frac{d\sigma_{11}}{dt} + \frac{\sigma_{11}}{\tau} \left(1 + \frac{\epsilon \tau}{\eta_0} (\sigma_{11} + \sigma_{22} + \sigma_{33})\right) - 2\dot{\gamma} \sigma_{12} &= 0 ,
\end{split} \\
\begin{split}
    \frac{d \sigma_{12}}{dt} + \frac{\sigma_{12}}{\tau} \left(1 + \frac{\epsilon \tau}{\eta_0} (\sigma_{11} + \sigma_{22} + \sigma_{33})\right) - \dot{\gamma} \sigma_{22} - \frac{\eta_0}{\tau} \dot{\gamma} &=0 ,
\end{split} \\
\begin{split}
    \frac{d \sigma_{22}}{dt} + \frac{\sigma_{22}}{\tau} \left(1 + \frac{\epsilon \tau}{\eta_0} (\sigma_{11} + \sigma_{22} + \sigma_{33})\right) &=0 ,
\end{split} \\
\begin{split}
    \frac{d \sigma_{33}}{dt} + \frac{\sigma_{33}}{\tau} \left(1 + \frac{\epsilon \tau}{\eta_0} (\sigma_{11} + \sigma_{22} + \sigma_{33})\right) &=0 .
\end{split} 
\end{align}
\end{subequations}

Because the $\sigma_{22}$ and $\sigma_{33}$ terms are factorizable from their respective equations, and we assume $\pmb{\sigma} = \pmb{0}$ at $t = 0$, they were set trivially to be zero. The first normal stress difference then becomes $N_1 = \sigma_{11}$.

The equations hence simplify to two coupled equations:
\begin{subequations} \label{linearPTT}
\begin{align}
\begin{split}
    \frac{d\sigma_{11}}{dt} + \frac{\sigma_{11}}{\tau} \left(1 + \frac{\epsilon \tau}{\eta_0} \sigma_{11} \right) - 2\dot{\gamma} \sigma_{12} &= 0 ,
\end{split} \\
\begin{split}
    \frac{d \sigma_{12}}{dt} + \frac{\sigma_{12}}{\tau} \left(1 + \frac{\epsilon \tau}{\eta_0} \sigma_{11}\right) - \frac{\eta_0}{\tau} \dot{\gamma} &=0 .
\end{split} 
\end{align}
\end{subequations}

We now proceed on the basis of a few physical arguments and substitute the strain rate function $\dot{\gamma}^e$ from Eq. \ref{eq:expshear}(b) into these evolution equations. We assume that at long times $\alpha t \gg 1$, the shear stress and the first normal stress difference scale with the instantaneous strain rate as $\sigma_{12} \sim [\dot{\gamma}^e]^x$ and $\sigma_{11} \sim [\dot{\gamma}^e]^y$, for constants $x$ and $y$ that we will find. We could have assumed a power scaling with the strain as well, but that is functionally the same since in exponential shear flow at long times, both the strain and strain rate are exponential functions and differ only by a pre-factor (cf. Eq. \ref{eq:expshear}). Any other lower order terms are negligible as a first approximation. 

These assumed functional forms can be substituted into Eq. \ref{linearPTT}(a) and (b) to produce the following two equations:
\begin{subequations} \label{linearPTT_scaling}
\begin{equation}
    y \alpha [\dot{\gamma}^e]^y + \frac{1}{\tau} [\dot{\gamma}^e]^y \left(1 + \frac{\epsilon \tau}{\eta_0} [\dot{\gamma}^e]^y\right) - 2[\dot{\gamma}^e]^{1+x} = 0 ,
\end{equation}
\begin{equation}
    x \alpha [\dot{\gamma}^e]^x + \frac{1}{\tau} [\dot{\gamma}^e]^x \left(1 + \frac{\epsilon \tau}{\eta_0} [\dot{\gamma}^e]^y \right) - \dot{\gamma}^e \frac{\eta_0}{\tau}=0 .
\end{equation}
\end{subequations}

The key to this analysis (and to the utility of exponential shear flow) is that the derivative of an exponential function is still an exponential function. In particular, the time derivative term, which is the first term in both equations, has the same exponent as the stress relaxation term (the second term in both equations). However, at large strain rates, the third term, which corresponds to the network softening term, is expected to dominate since this higher-order effect is nonlinear, and this must be balanced by the last term, which is the forcing function arising from the applied strain rate. The ``method of dominant balance'' indicates that only the last two terms of each equation are dominant, and hence identifies that the time derivative term is negligible. By equating the exponents of these two terms, we find that $2y = 1 + x$ from Eq. \ref{linearPTT_scaling}(a) and $x + y = 1$ from Eq. \ref{linearPTT_scaling}(b), which can be solved simultaneously to obtain $x = 1/3$ and $y = 2/3$. These are in agreement with the exponents of the leading order terms for the strain rate as calculated by Shogin \cite{shogin_full_2021} and which give Eq. \ref{shogin} above.

We can further verify this by neglecting all terms apart from the last two in Eq. \ref{linearPTT}, leaving the following two algebraic equations:
\begin{subequations}
\begin{align}
\begin{split}
    \frac{\sigma_{11}}{\tau} \frac{\epsilon \tau}{\eta_0} \sigma_{11} - 2\dot{\gamma} \sigma_{12} &= 0 ,
\end{split} \\
\begin{split}
    \frac{\sigma_{12}}{\tau} \frac{\epsilon \tau}{\eta_0} \sigma_{11} - \frac{\eta_0}{\tau} \dot{\gamma} &=0 .
\end{split} 
\end{align}
\end{subequations}

Solving these equations simultaneously, we obtain exactly the first order analytical expressions for the shear stress and $N_1$ calculated by Shogin \cite{shogin_full_2021} in Eq. \ref{shogin}, complete with accurate pre-factors.

\begin{figure*}
\centering
\includegraphics[width=0.9\textwidth]{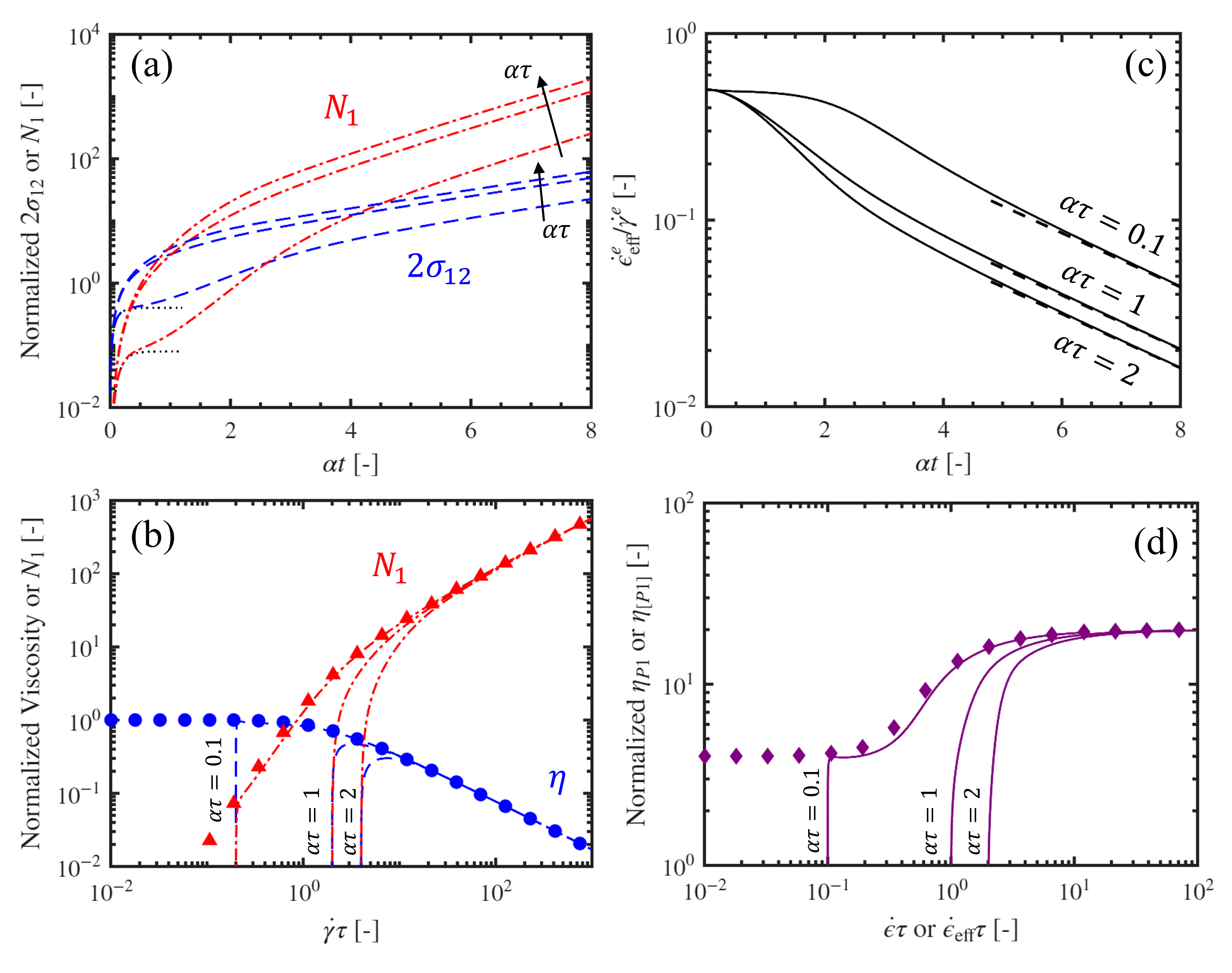}
\caption{Exponential shear flow of the linear Phan-Thien--Tanner (PTT) model with network softening parameter $\epsilon = 0.1$ for different dimensionless flow strengths ($\alpha\tau = 0.1, 1, 2$). All stress quantities are normalized by a modulus $G = \eta_0 / \tau$, viscosities are normalized by the shear viscosity $\eta_0$, and shear and extension rates by $1/\tau$. All symbols correspond to steady state data and lines show the trajectory of material responses computed from the time-evolving shear stress and $N_1$ measured in exponential shear flow. (a) The shear stress and first normal stress difference against dimensionless time. Curves of increasing $\alpha \tau$ are arranged from bottom to top. The leading order responses to an initial step strain rate for $\alpha \tau = 0.1$ are indicated by black dotted lines for both the shear stress and $N_1$. (b) Evolution of the steady shear viscosity and first normal stress difference as a function of the dimensionless strain rate. (c) Evolution of the normalized effective extension rate against dimensionless time. Curves of increasing $\alpha \tau$ are arranged from left to right. (d) The composite planar extensional viscosity function against effective extension rate, which all plateau to the same value at large extension rates. The diamond symbols correspond to true planar extensional viscosities computed analytically. Curves of increasing $\alpha \tau$ are labeled from left to right.} 
\label{fig:linearptt}
\end{figure*}

To validate this asymptotic analysis, we simulate the linear PTT model subject to exponential shear with different $\alpha$ values. The stress responses are shown in Fig. \ref{fig:linearptt}(a). We show in Fig. \ref{fig:linearptt}(b) that for various chosen values of $\alpha$, we now get better agreement with the steady shear material functions at long times for all values of $\alpha$ with no offset factors of the form $(1 + \alpha \tau)$, as expected from the ``method of dominant balances''. We next plot the dimensionless effective extension rate normalized by the instantaneous shear rate in Fig. \ref{fig:linearptt}(c). As before, this starts at a value of $\dot{\epsilon}_\mathrm{eff}^e / \dot{\gamma}^e (t = 0) = 0.5$ and then (in contrast to our analysis of the UCM model) monotonically decreases with a power law of exponent $-1/3$. This is because the value of $\dot{\epsilon}_\mathrm{eff}^e / \dot{\gamma}^e \rightarrow \sigma_{12} /N_1 \simeq (4\tau / \epsilon)^{-1/3} [\dot{\gamma}^e]^{-1/3}$ at large strain rates, and the leading power law is the same as for the ratio of the two steady shear material functions.

The ``method of dominant balances'' is thus a systematic way of identifying the terms that can be neglected in the asymptotic limit of $\alpha t \gg 1$ under exponential shear. This approach is generally valid for any nonlinear constitutive model with the UCM model as its backbone, because other nonlinear mechanisms will similarly dominate at high strain rates, thereby leading to a dominant balance that excludes the time derivative term $\tfrac{d \pmb{\sigma}}{dt}$ and the present value of stress $\pmb{\sigma}$. For the UCM model, because there are no other nonlinear terms, the dominant balance between the stress relaxation term and the strain rate forcing term leads to unwanted pre-factors of $\alpha \tau$ that modify the material functions defined in Eq. \ref{expviscosities} and \ref{compositematerialfunction}. In contrast, any other nonlinear model (such as the linear PTT model discussed in this section) will have minimal dependence on $\alpha$ for the material functions in the limit of $\alpha t \gg 1$. Further analysis of another popular nonlinear model, the Giesekus model \cite{giesekus_simple_1982}, following the arguments by Khair \cite{khair_large_2016} for large amplitude oscillatory shear, is discussed in Appendix A. It may be more difficult to identify analytically the dominant balance for more microscopically-motivated models that incorporate multiple relaxation mechanisms. For example, the Rolie-Poly model \cite{likhtman_simple_2003} for entangled polymer melts includes reptation, chain stretch and convective constraint release. However, we show through numerical simulations in Appendix B that exponential shear is also able to replicate the steady state material functions.

\section{Representative Experimental Data}

Two benchmark materials are chosen to illustrate the utility of exponential shear as a flow protocol. The first is a polymer melt, polydimethylsiloxane (PDMS), which is often used as a calibration standard in rheometers. The second is a weakly entangled polyisobutylene (PIB) solution, which is representative of materials that are widely used in industry as adhesives, coolants, and lubricants. Such solutions are not viscous enough to characterize in extension using rotating clamp devices \cite{sentmanat_miniature_2004}, yet are too viscous to use in microfluidic setups such as OSCER \cite{haward_optimized_2012}. PDMS has a relaxation spectrum that can be compactly described with a few relaxation modes, while the PIB solution has a broad relaxation spectrum well-described by Zimm dynamics.

\subsection{Materials}

The polydimethylsiloxane (PDMS) fluid is ``BELSIL\textregistered~DM 1000000 PDMS'', a linear unmodified PDMS purchased from the company Wacker. It has a zero-shear viscosity of approximately 1000 Pa.s at $25^\circ$C. The 2 wt\% polyisobutylene (PIB) solution in paraffin oil has been well studied by More \textit{et al.} \cite{more_elasto-inertial_2024}. The dissolved polymer has a weight-averaged molecular weight of about $10^6$ g/mol and a zero-shear viscosity of approximately 5 Pa.s at $20^\circ$C. All experiments were performed on an ARES-G2 rheometer. For the PDMS fluid, a stainless steel cone (0.04 rad 25 mm) from TA Instruments was used. For the PIB solution, a stainless steel cone (0.1 rad 25 mm) from TA Instruments was used for oscillatory shear and steady shear measurements, while a stainless steel 2$^\circ$ 40 mm cone from TA Instruments was used for exponential shear measurements. For the 25 mm cones, the truncation height at the virtual apex of the cone was set as 50 $\mu$m, while for the 40 mm cone, it was 45 $\mu$m. 

\subsection{Small Amplitude Oscillatory Shear}

\begin{figure*}
\centering
\includegraphics[width=0.5\textwidth]{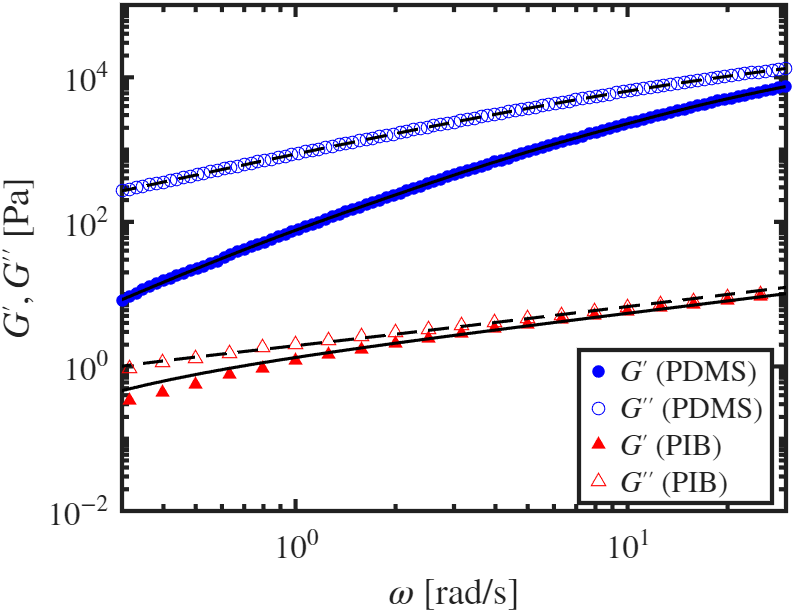}
\caption{Small amplitude oscillatory shear for the PDMS entangled polymer melt sample at a temperature of 25$^\circ$C and the 2 wt\% PIB solution at 20$^\circ$C. Experimental data are shown as symbols and the black lines are fits to the data using a generalized discrete Maxwell spectrum and the Zimm model respectively.} 
\label{fig:saos}
\end{figure*}

Small Amplitude Oscillatory Shear tests were first performed to determine the characteristic relaxation time for each test fluid. A frequency sweep was performed at 10 \% strain (i.e. $\gamma_0 = 0.1)$ and 25$^\circ$C for the PDMS, and at 1\% strain (i.e. $\gamma_0 = 0.01$) at 20$^\circ$C for the PIB solution. We show the resulting SAOS data for both materials in Fig. \ref{fig:saos}.

\begin{table*}
\caption{Fits of the experimental data to a multi-mode generalized discrete Maxwell spectrum for the PDMS fluid, with a zero-shear viscosity $\eta_0^{PDMS} = 897$ Pa.s at a temperature of $25^\circ$C, and viscosity-weighted relaxation time $\bar{\tau}_\eta = 0.107$ s and polydispersity \textit{\DJ} = 6.70.}
\begin{ruledtabular}
\begin{tabular}{ccc} \label{pdmssaos}
Mode $i$ & Modulus $G_i$ [Pa] & Relaxation Time $\tau_i$ [s] \\
\hline
1 & $5.80 \times 10^1$ & $8.60 \times 10^{-1}$ \\
2 & $8.43 \times 10^2$ & $2.02 \times 10^{-1}$ \\
3 & $6.15 \times 10^3$ & $5.13 \times 10^{-2}$ \\
4 & $4.91 \times 10^4$ & $7.34 \times 10^{-3}$ \\
\end{tabular}
\end{ruledtabular}
\end{table*}

The SAOS data from the PDMS sample can be fitted with a 4-mode discrete relaxation spectrum and the parameters are in Table \ref{pdmssaos}. This produces a viscosity-weighted relaxation time (Eq. \ref{viscosityweightedrelaxationtime}) of $\bar{\tau}_\eta = 0.107$ s. The modulus-weighted relaxation time is $\bar{\tau}_G = (\sum_{i=1}^N G_i \tau_i) / (\sum_{i=1}^N G_i) = 0.0160$ s. The polydispersity \textit{\DJ} is hence $\bar{\tau}_\eta / \bar{\tau}_G = 6.70$.

The SAOS data for the PIB solution was obtained by More \textit{et al.} \cite{more_elasto-inertial_2024} over a wider range of frequencies using time-temperature superposition and we have confirmed that these values agree with the current sample within the window of frequencies probed in this work. More \textit{et al.} \cite{more_elasto-inertial_2024} fitted the SAOS data to the following two expressions for the Zimm model provided by Rubinstein and Colby \cite{rubinstein_polymer_2003}: 
\begin{subequations}
\begin{equation}
    G'(\omega) = G_c \frac{\omega \tau_Z \sin(\chi ~\text{atan}(\omega \tau_Z))}{[1 + (\omega \tau_Z)^2]^{\chi/2}}
\end{equation}
\begin{equation}
    G'' (\omega) = G_c \frac{\omega \tau_Z \cos(\chi ~\text{atan}(\omega \tau_Z))}{[1 + (\omega \tau_Z)^2]^{\chi/2}}
\end{equation}
\end{subequations}

with three fitted parameters: a characteristic modulus $G_c = 0.96$ Pa, a Zimm relaxation time $\tau_Z = 4.99$ s, and $\chi = 1 - 1/(3\nu)$, in which the solvent quality $\nu = 0.593$. The Zimm spectrum consists of an infinite number of relaxation modes that are weighted equally, with the following relation between the relaxation time of each $i^\text{th}$ mode and the Zimm relaxation time: 
\begin{equation}
    \tau_i = \frac{\tau_Z}{i^{3\nu}}
\end{equation}
Using the Riemann zeta function \cite{rubinstein_polymer_2003}, the viscosity-weighted relaxation time for the Zimm spectrum of the PIB solution is:
\begin{equation}
    \bar{\tau}_\eta = \frac{\sum_{i=1}^n \eta_i \tau_i}{\sum_{i=1}^n \eta_i} = \frac{\sum_{i=1}^n \tau_i^2}{\sum_{i=1}^n \tau_i} = \tau_Z \frac{\sum_{i=1}^n \frac{1}{i^{6\nu}}}{\sum_{i=1}^n \frac{1}{i^{3\nu}}} = \tau_Z \frac{\zeta(6\nu)}{\zeta(3\nu)} = 2.92 ~\text{s}
\end{equation}

In Appendix C, we show that the entangled PDMS also follows the empirical Cox-Merz rule \cite{cox_correlation_1958} and Laun's rule \cite{laun_prediction_1986,das_launs_2024}, but the PIB solution does not. Although many other classes of complex fluids do not  follow these correlations, if a material such as a well-entangled homopolymer melt or solution  does follow these rules, an \textit{estimate} of the steady-state flow curve and the steady planar extensional viscosity can be obtained even without using exponential shear by combining these rules with Eqs. \ref{compositematerialfunction}(a,b) . 

\subsection{PDMS Melt}

\begin{figure*}
\centering
\includegraphics[width=\textwidth]{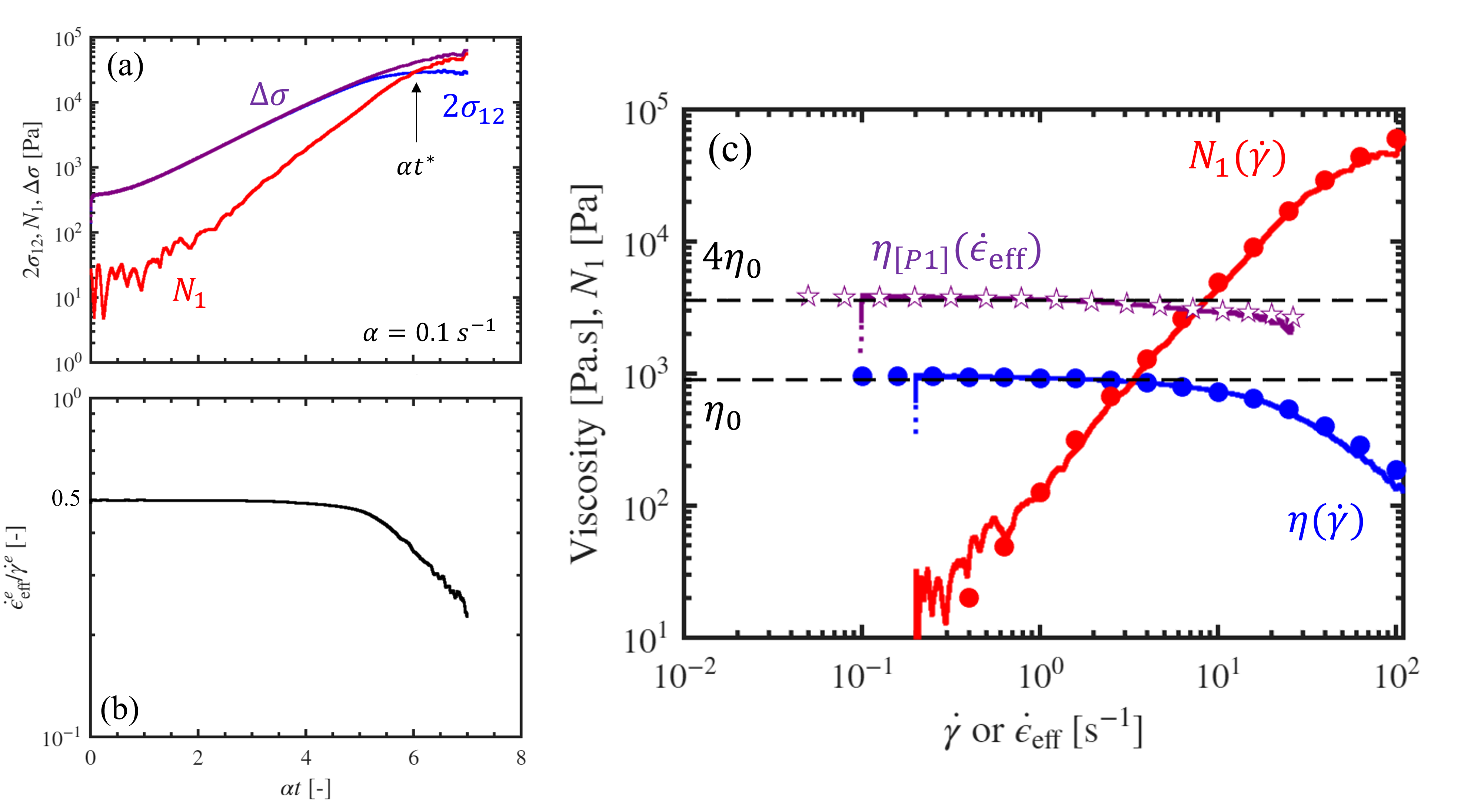}
\caption{Exponential shear for PDMS at $\alpha = 0.1$ s$^{-1}$, with $\alpha \bar{\tau}_\eta = 0.0107 \ll 1$, and for a final time $t_f = 70$ s. All large symbols are steady state data obtained from independent rate sweep experiments, and smaller continuous dots represent measurements from time-varying exponential shear $\dot{\gamma}^e$, starting from $\dot{\gamma}^e_0 = 0.2$ s$^{-1}$ to $\dot{\gamma}^e_f = 100$ s$^{-1}$. (a) The shear stress $\sigma_{12}$, first normal stress difference $N_1$, and principal stress difference $\Delta \sigma$ against dimensionless time. (b) Evolution of the normalized effective extension rate against dimensionless time. (c) The steady shear viscosity and first normal stress difference against shear rate, together with the composite planar extensional viscosity function shown as a function of the effective extension rate (Eq. \ref{compositematerialfunction}). The shear stress data is the raw data while the $N_1$ data was filtered using a moving average filter with a window width of 50 samples.} 
\label{fig:pdms}
\end{figure*}

We subject both the PDMS sample and the PIB solution to exponential shear by prescribing a strain $\gamma(t) = 2 \sinh(\alpha t)$ using the ``Arbitrary Wave'' function on the ARES-G2 rheometer and setting $\alpha = 0.1$ s$^{-1}$. This choice of $\alpha$ satisfies the criterion that $\alpha \bar{\tau}_\eta \ll 1$ for both materials such that the fluid response is quasisteady and therefore accurate approximations for the steady shear viscosity and planar extensional viscosity can be obtained. We now analyze the data generated from exponential shear experiments using the proposed material functions (Eq. \ref{expviscosities} and \ref{compositematerialfunction}). Because edge fracture is known to occur in the PDMS sample at high strain rates, exponential shear at $\alpha = 0.1$ s$^{-1}$ ($\alpha \bar{\tau}_\eta = 0.0107$) was performed for 70 s, corresponding to a Hencky strain $\alpha t_{max} \simeq 7$ and a maximum strain rate of $100$ s$^{-1}$ ($Wi_{max} = \bar{\tau}_\eta \dot{\gamma}_{max} = 10.7$). As we show in Fig \ref{fig:pdms}(a), there is an exponential increase in both the shear stress and the first normal stress difference. The point at which $2\sigma_{12}$ and $N_1$ cross over is $\alpha t^* \approx 6$. This crossover occurs at relatively high values of $\alpha t$, indicating that the sample is only weakly elastic. Hence, the principal stress difference $\Delta \sigma$ is dominated by the shear stress contribution until large values of $\alpha t$. We illustrate how the effective extension rate changes with $\alpha t$ in Fig \ref{fig:pdms}(b). It remains near $\dot{\epsilon}_\mathrm{eff}^e / \dot{\gamma}^e = 0.5$ at the start (cf. Section III) and rapidly decreases beyond a dimensionless time of $\alpha t \approx 5$ as the normal stress difference $N_1 (t)$ begins to grow and ultimately exceed $2\sigma_{12} (t)$.

In Fig. \ref{fig:pdms}(c), the steady state shear viscosity $\eta(\dot{\gamma})$ and $N_1 (\dot{\gamma})$ obtained from an independent shear rate sweep over the range $0.1 \leq \dot{\gamma} \leq 100$ s$^{-1}$ are compared directly with the material functions obtained from exponential shear (Eq. \ref{expviscosities}). After a very short initial transient response to the imposed deformation rate  $\dot{\gamma}^e_0 = 0.2$ s$^{-1}$, we see good agreement even up to high strain rates. This confirms that for a real material, which exhibits a spectrum of relaxation times rather than a single distinct one, exponential shear passes through a series of quasi-steady states as long as $\alpha \bar{\tau}_\eta \ll 1$. 

Using Eq. \ref{compositematerialfunction}, we can also calculate an approximation to the steady planar extensional viscosity. This is compared with the same material function calculated using only steady shear data, which was first proposed by King and McKinley \cite{king_kinematic_2026} and represented as hollow stars in Fig. \ref{fig:pdms}(c). There is good agreement between the two estimates (shown by purple dots and the hollow purple star symbols respectively), which is unsurprising given the close agreement between the exponential shear and steady shear data. At low strain rates, the calculated composite material function is exactly four times the zero-shear viscosity, in agreement with the Newtonian limit of the Trouton ratio in planar extension. This material function subsequently shows weak extensional softening behavior at higher extension rates. For a linear polydisperse entangled polymer network such as PDMS, there is a lack of microscopic features such as long chain branching that can induce strong polymer stretch. At higher extension rates, polymer chains start to orient and pose less resistance to the flow, leading to extensional softening behavior. 

\subsection{2 wt\% Polyisobutylene Solution in Paraffin Oil}

\begin{figure*}
\centering
\includegraphics[width=\textwidth]{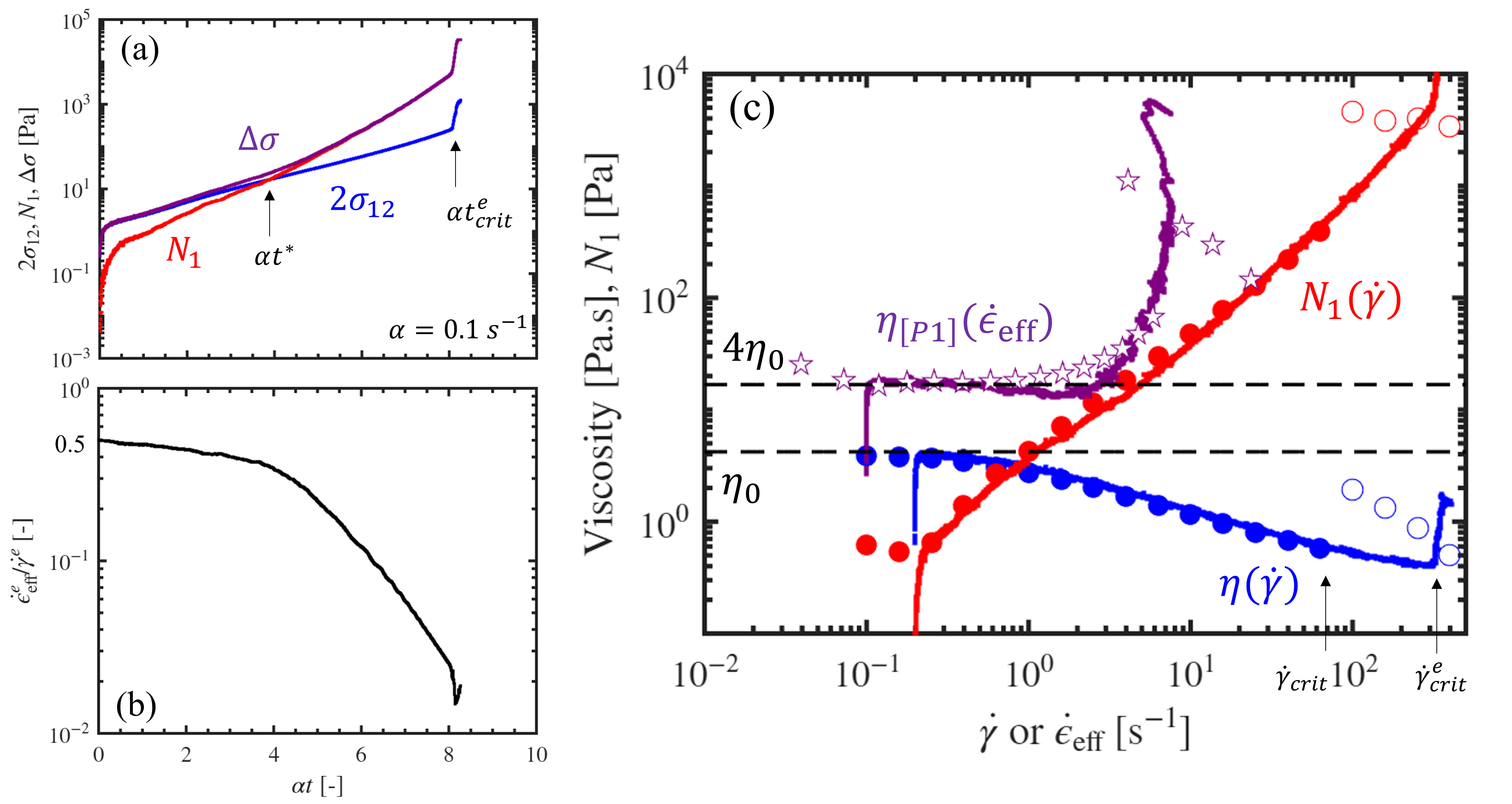}
\caption{Exponential shear for the PIB solution at $\alpha = 0.1$ s$^{-1}$, with $\alpha \bar{\tau}_\eta = 0.292 \ll 1$, and for a final time $t_f = 85.2$ s. All large symbols are steady state data and smaller continuous dots represent measurements from time-varying exponential shear $\dot{\gamma}^e$, starting from $\dot{\gamma}^e_0 = 0.2$ s$^{-1}$ to $\dot{\gamma}^e_f = 500$ s$^{-1}$. (a) The shear stress $\sigma_{12}$, first normal stress difference $N_1$, and principal stress difference $\Delta \sigma$ against dimensionless time. (b) Evolution of the normalized effective extension rate against dimensionless time. (c) The steady shear viscosity and first normal stress difference against strain rate, together with the composite planar extensional viscosity function against effective extension rate. The exponential shear viscosity function (Eq. \ref{expviscosity}) is calculated from raw shear stress data while the $N_1$ data was filtered using a moving average filter with a window width of 50 samples. Large hollow symbols correspond to the onset of an elastic instability beyond $\dot{\gamma}_{crit} = 80$ s$^{-1}$ and are only shown for comparative purposes.} 
\label{fig:pib}
\end{figure*}

We next analyze the data for the 2 wt\% PIB solution. This is a semi-dilute polymer solution, as characterized by More \textit{et al.} \cite{more_elasto-inertial_2024}. We performed exponential shear on the sample at $\alpha = 0.1$ s$^{-1}$ ($\alpha \bar{\tau}_\eta = 0.292$) for 85.2 s, corresponding to a maximum strain rate of $500$ s$^{-1}$ ($Wi_{max} = \bar{\tau}_\eta \dot{\gamma}_{max} = 1460$). As we show in Fig. \ref{fig:pib}(a), there is an exponential increase in both the shear stress and the first normal stress difference, up to a critical time $\alpha t_{crit}^e \approx 8$, corresponding to a shear rate of $\dot{\gamma}^e_{crit} \approx 300$ s$^{-1}$. The PIB solution is also relatively more elastic than the PDMS sample, evidenced by noting that  the point at which $2\sigma_{12}$ and $N_1$ cross over occurs at $\alpha t^* \approx 4$. In Fig. \ref{fig:pib}(b), the ratio of the effective extension rate to the strain rate also decreases much earlier in time from its initial value of $\dot{\epsilon}_\mathrm{eff}^e / \dot{\gamma}^e = 0.5$, compared to the PDMS sample in Fig. \ref{fig:pdms}. Beyond the critical point $\dot{\gamma}^e_{crit} = 300$ s$^{-1}$, there is a sudden drastic increase in both the shear stress and $N_1$. Video imaging shows that this corresponds to the onset of an elasto-inertial instability which ultimately leads to the fluid being ejected from the gap and subsequently climbing up the cone. All measurements at higher strain rates than this critical strain rate $\dot{\gamma}^e_{crit}$ are hence unreliable. 

In Fig. \ref{fig:pib}(c), we see good agreement between the steady shear material functions and the material functions calculated from exponential shear (Eq. \ref{expviscosities}). Remarkably, even though the steady shear measurements become unreliable beyond $\dot{\gamma}_{crit} \approx 80$ s$^{-1}$ ($Wi = 234$), the exponential shear protocol is able to temporarily suppress the elasto-inertial instability up to higher rates $\dot{\gamma}^e_{crit} \approx 300$ s$^{-1}$ ($Wi = 876$). This difference may arise because the steady shear measurements were performed with a 25 mm cone rather than a 40 mm cone, and from the dynamic time-varying nature of the exponential shear protocol. 

An approximation to the steady planar extensional viscosity using Eq. \ref{compositematerialfunction} (purple closely spaced dots in Fig. \ref{fig:pib}(c)) shows good agreement with the same material function calculated using only the steady shear data (purple hollow stars in Fig. \ref{fig:pib}(c)). The curve for the steady planar extensional viscosity $\eta_{[P1]}$ (purple closely spaced dots in Fig. \ref{fig:pib}(c)) starts at four times the shear viscosity at low strain rates, and subsequently increases drastically at a critical effective extension rate $\dot{\epsilon}_{CS} \approx 5$ s$^{-1}$. This is indicative of the well-known coil-stretch transition \cite{de_gennes_coil-stretch_1974} for polymer solutions and the increasing tensile stress difference arises from the enhanced resistance posed by the stretching of the long polymer chains in solution. Notably, beyond the critical shear rate for the elasto-inertial instability, $\dot{\gamma}^e_{crit} = 300$ s$^{-1}$, which corresponds to a critical extension rate $\dot{\epsilon}_{crit} \approx 7$ s$^{-1}$, the curve for the steady planar extensional viscosity $\eta_{[P1]}$ bends backward, becoming multi-valued at a single effective extension rate. This is likely an artifact of the instability and again indicates that the data beyond that strain rate are unreliable.

\section{Experimental Considerations and Operating Windows}

\begin{figure*}
\centering
\includegraphics[width=\textwidth]{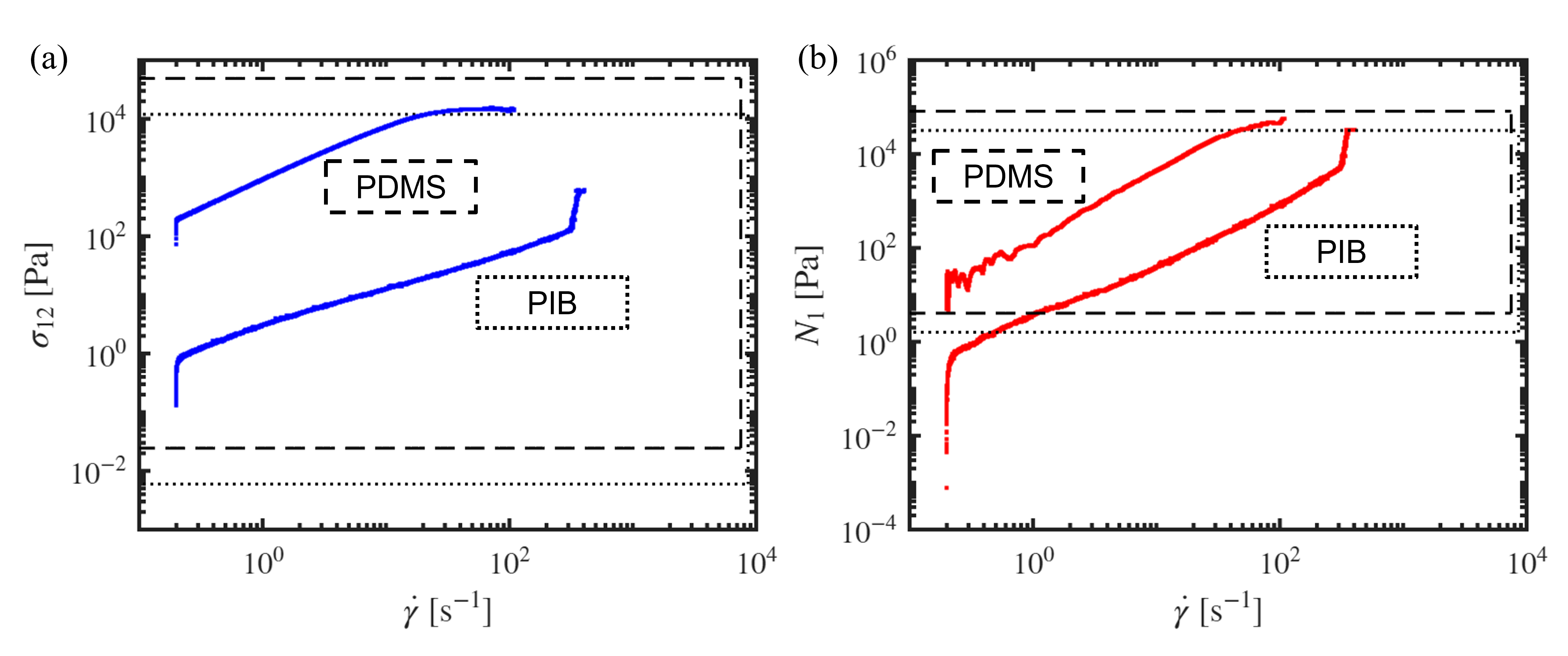}
\caption{Operating windows for (a) the measured shear stress $\sigma_{12}(t)$ and (b) the measured first normal stress difference $N_1(t)$ for the PDMS melt and the PIB solution at $\alpha = 0.1$ s$^{-1}$. The dashed lines represent the operating windows for the geometry used for the PDMS sample (0.04 rad 25 mm cone and plate), while the dotted lines represent the operating windows for the geometry used for the PIB solution (2$^\circ$ 40 mm cone and plate).} 
\label{fig:operatingwindow}
\end{figure*}

Through analysis of two fluids, a polymer solution (PIB) and melt (PDMS), we have illustrated how exponential shear can be used to rapidly obtain information about both the shear and extensional rheological properties of a material. However, there are several experimental constraints to consider. Exponential shear is a severely nonlinear test since the shear rate increases exponentially, hence the accessible operating window of an exponential shear experiment is constrained by the operating limits of the rheometer and the geometry used. Inspired by the earlier presentations of Pipe \textit{et al.} \cite{pipe_high_2008} for microfluidic devices, Ewoldt \textit{et al.} \cite{spagnolie_experimental_2015} for rheometry in general, and Warwaruk \textit{et al.} \cite{warwaruk_dripping-onto-substrate_2026} for dripping-onto-substrate experiments, we construct an operating window for exponential shear in terms of the shear rate applied and the stresses measured \cite{kroo_unifying_2025}. These are constrained by the manufacturer specifications for the rheometer. According to TA Instruments, the torque range of the ARES-G2 is $M = 0.1$ $\mu$N.m to 200 mN.m under steady shear, while the normal force range is $F = 0.001$ to 20 N. The angular velocity range is $\Omega =1 \times 10^{-6}$ rad/s to 300 rad/s. We use these values to construct bounds on the operating space as shown in Fig. \ref{fig:operatingwindow}.

For clarity, the shear stress and the $N_1$ measured for each sample in Section V are plotted on separate graphs in Fig. \ref{fig:operatingwindow}. In both figures, the vertical dotted line represents the maximum shear rate attainable by the rheometer, which is related to the maximum angular velocity $\Omega_\mathrm{max}$ and the cone angle $\beta$:
\begin{equation}
    \dot{\gamma}_\textrm{max} = \frac{\Omega_\mathrm{max}}{\beta}
\end{equation}

For Fig. \ref{fig:operatingwindow}(a), the horizontal lines correspond, respectively, to the maximum and minimum shear stress that are experimentally attainable. These are related to the maximum torque $M_\mathrm{max}$ and minimum torque $M_\mathrm{min}$ measurable by the rheometer and the cone plate diameter $D$ by the expressions:
\begin{subequations}
\begin{equation}
    \sigma_{12, \mathrm{min}} = \frac{3 M_\mathrm{min}}{2\pi (D/2)^3}
\end{equation}
\begin{equation}
    \sigma_{12, \mathrm{max}} = \frac{3 M_\mathrm{max}}{2\pi (D/2)^3}
\end{equation}
\end{subequations}

Similarly for Fig. \ref{fig:operatingwindow}(b), the horizontal lines correspond to the maximum and minimum first normal stress difference that are experimentally attainable. These are related to the maximum axial force $F_\mathrm{max}$ and minimum axial force $F_\mathrm{min}$ measurable by the rheometer and the cone plate diameter $D$:
\begin{subequations} \label{n1bound}
\begin{equation}
    N_{1,\mathrm{min}} = \frac{2 F_\mathrm{min}}{\pi (D/2)^2}
\end{equation}
\begin{equation}
    N_{1,\mathrm{max}} = \frac{2 F_\mathrm{max}}{\pi (D/2)^2}
\end{equation}
\end{subequations}

In practice, other instabilities can also compromise data fidelity at high shear rates. As noted, for the PIB solution the onset of an elasto-inertial instability at high deformation rates leads to a sudden and drastic increase in the shear stress and $N_1$. This can be seen clearly in both Fig. \ref{fig:operatingwindow}(a) and (b) at a strain rate of about 300 s$^{-1}$. Due to the more limited dynamic range of the normal force transducer, it is expected that this  provides the more stringent limit on successful implementation of exponential shear experiments. Indeed, for the PIB sample, the sudden increase in $N_1$ at a critical strain rate caused the limit of the maximum force measurable to be exceeded, and the arbitrary wave protocol was automatically stopped by the rheometer software. Edge fracture is another possible failure mode, and a strong indication of this is if the stress starts to rapidly decrease as the shear rate is increased. This indicates that the velocity field is no longer homogeneous and that subsequent dynamical measurements are unreliable. 
In both cases, the critical shear rate for the onset of instability is material-dependent and difficult to predict \textit{a priori}, but an additional benefit of performing an exponential shear experiment with a time-varying shear rate profile of the form given in Eq. \ref{eq:expshear}(b) is that it can offer a quick estimate of the maximum shear rate beyond which measurements become unreliable compared to performing a long sequence of steady shear experiments at discrete shear rates. 

Obtaining information about the extensional viscosity of the material requires accurate measurements of the first normal stress difference. This is typically challenging to access for several reasons \cite{costanzo_review_2024}. For one, accurate normal force measurements are highly sensitive to the shape of the free surface at the edge of the sample, requiring careful sample trimming. Venerus \cite{venerus_free_2007} has calculated that distorted free surfaces can lead to deviations of 10\% or more from the true value of $N_1$.  Because Eq. \ref{compositematerialfunction} relies on a nonlinear combination of the shear rate, shear stress, and $N_1$, any experimental errors in the measurement of $N_1$ are greatly amplified. Furthermore, it is known that the axial compliance of the normal force transducer in a rheometer can adversely affect measurements for transient experiments \cite{hansen_transient_1975,schweizer_role_2006}. This is due to the generation of a (reverse) squeeze flow, and is another reason to choose a small value of $\alpha$, such that initial transients (which arise from the linear viscoelastic response to a step strain rate) do not affect the measurements of interest at $\alpha t \gg 1$. 

The exponential shear protocol we have outlined in this work is especially well suited to viscous polymer solutions and moderately entangled polymer melts. For dilute polymer solutions or surfactant solutions with shear viscosity close to that of water, extensive studies of the extensional response have been carried out over the last two decades using filament thinning \cite{mckinley_filament-stretching_2002} or microfluidic \cite{haward_optimized_2012} devices. However, the major limitation of performing exponential shear on such mobile fluid systems is that the values of $N_1$ generated are very low. Successful application of exponential shear depends on a measurable $N_1$ signal in order to calculate extensional properties. Even if $N_1$ is one or two orders of magnitude higher than the shear stress at sufficiently high strain rates, the generated values of $N_1$ could still be below the noise floor of typical separate-motor-transducer (SMT) rheometers.  Recent developments in simultaneous measurement of birefringence and stress on a rheometer, also known as the `Rheo-Iris' \cite{araida_analysis_2025}, may offer an alternate route in the future to normal stress difference determination in dilute solutions and weakly viscoelastic gels through measurement of the shear-induced optical retardance and application of the stress-optic rule. 

The diagonal trajectories followed by the PDMS and PIB data shown in Fig. \ref{fig:operatingwindow}(b) make maximal use of the dynamic range of the ARES normal force transducer. Selecting different values of the test fixture diameter $D$ shift the minimum and maximum detectable values of $N_1$ up or down respectively according to Eq. \ref{n1bound}. Similarly, changing the value of $\alpha$ imposed in the experiment shifts the initial value of the shear rate imposed ($\dot{\gamma}^e_0 = 2 \alpha$) and also changes the slope of the (exponentially-varying) trajectory of the normal stress difference (cf. Eq. 9b and 28b) through the operating space. Judicious choice of test fixture and exponential rate parameter $\alpha$ is thus essential, but with care, a single exponential shear flow experiment can enable the full flow curves of $\eta(\dot{\gamma})$, $\Psi_1(\dot{\gamma})$ and $\eta_{[P1]}(\dot{\epsilon}_\mathrm{eff})$ to be mapped out in a single experimental test lasting about 100 s.

\section{Conclusion}

We have shown that exponential shear flow with a profile for $\dot{\gamma}^e (t)$ given by Eq. \ref{eq:expshear} can rapidly sweep out an entire flow curve within a single test for carefully chosen values of $\alpha$. Furthermore, by recording both the time-evolving torque and the normal force in a rheometer, we can not only obtain the steady shear viscosity function $\eta (\dot{\gamma})$ and the first normal stress difference $N_1 (\dot{\gamma})$, but also the new composite material function $\eta_{[P1]} (\dot{\epsilon}_\mathrm{eff})$ which we have shown provides a good estimate of the steady planar extensional viscosity function of the fluid. The proposed material functions (Eq. \ref{expviscosities} and \ref{compositematerialfunction}) thus hopefully resolve the debate surrounding the definition of appropriate material functions for analyzing exponential shear experiments. 

To support the use of these material functions, we have presented a detailed analysis of the kinematics of exponential shear flow.
% and proposed material functions that best represent both the shear and extensional characteristics of the flow. 
We agree with the conclusion of Venerus and Schieber \cite{venerus_exponential_2000,neergaard_exponential_2000} that the kinematic history of a material element under exponential shear is different from that undergoing transient planar extension. However, this is not due to the presence of vorticity in the flow, which can be accounted for by taking the appropriate projection of the strain rate tensor onto the eigenvectors of the stress tensor \cite{king_kinematic_2026}. Instead, we show that the time-varying kinematics of exponential shear flow are fundamentally different from both steady shear and planar extension, and this can be clearly visualized in our map of kinematic trajectories presented in Fig. \ref{fig:invariantmap}(c). Appropriate definitions of the time-varying Deborah number as $De(t) = \alpha \tau \coth(\alpha t)$ and the time-varying Weissenberg number as $Wi(t) = \tau \dot{\gamma} = 2 \alpha \tau \cosh(\alpha t)$ provide measures of the flow unsteadiness and the flow strength respectively. Projecting these trajectories of exponential shear onto a Pipkin diagram \cite{ewoldt_mapping_2017} shows that they move from the lower right (representing an initial unsteady linear viscoelastic response) to the upper left, towards the region that represents nonlinear steady state viscometric flows.

Almost 60 years ago, Metzner and Metzner \cite{metzner_stress_1970} noted that ``experimentally, extensional flows are difficult to generate under well-defined conditions and the available studies are subject to important restrictions as to the deformation rate levels attained or the duration of the deformation process, or are studies in which
the analysis of the deformation process was
only indirect or approximate''. The natural question to ask is whether exponential shear or, more precisely, the kinematic equivalence between steady shearing and planar extensional flows, can supplant existing techniques for measuring the extensional viscosity of fluids. Several authors \cite{graham_using_2001,suneel_characterisation_2003,wagner_exponential_2005} have made the point that exponential shear is not a sufficiently stringent test of the constitutive equation of an unknown viscoelastic fluid. We can assess this claim using the maps of invariants presented in Fig. \ref{fig:invariantmap}. From a purely kinematic point of view, exponential shear trajectories move along the same line that steady shear or planar extension traverses. A more accurate statement should thus be that shearing flows, planar extension and exponential shear produce the same rheological information and are not by themselves sufficiently stringent tests for uniquely identifying a constitutive equation of an unknown complex fluid. Uniaxial or biaxial extensional measurements also access the third invariant of the strain rate tensor (i.e., deformations with $III_{\pmb{\dot{\gamma}}} \neq 0$). 
These flows lie off of the line corresponding to all planar flows and probe a different part of the full three-dimensional kinematic space, hence such protocols are expected to be more stringent tests of a material's constitutive behavior when combined with shear measurements. Filament stretching or microfluidic devices that generate or approximate uniaxial extension or biaxial compression will therefore continue to play a crucial role in the characterization of the extensional properties of complex fluids \cite{mckinley_filament-stretching_2002,petrie_one_2006,malkin_polymer_2014}. 

We envision that exponential shear tests of the kind described here will be a useful technique for rapidly characterizing the nonlinear properties of a range of viscoelastic fluids. However, several criteria must be met for accurate results to be obtained. The value of $\alpha$ needs to be carefully chosen such that $\alpha \bar{\tau}_\eta \ll 1$ to enable a quasisteady state to be rapidly achieved. Both the time-varying torque and normal force must be measured in the experiment. The exponential shear flow experiment should also be run for a sufficiently long time to reach high strain rates, $Wi \gg 1$. This final time is limited by the maximum imposable shear rate of the rheometer, and by the maximum torque and normal force limits given by the operating windows calculated in Section VI. As we showed in our measurements for the PIB solution, we can also ascertain the onset of torsional (curved-streamline) flow instability within a single experiment and at much faster experimental times for an appropriately chosen value of $\alpha$. This technique may thus be particularly useful for strongly elastic liquids such as polymer melts as it seems able to delay the onset of edge fracture at large deformations.

We are also testing the validity of the relations given in Eq. \ref{compositematerialfunction} in a wider class of complex fluids that also exhibit yield stresses, and it is expected that once the imposed deformation moves beyond the yield strain, the rate-thinning behavior of the material and its dependence on flow type can similarly be measured using exponential shear. Finally, future work will focus on using this rapid strongly nonlinear flow as a means of rapidly generating the large volumes of high-fidelity rheometric training and test data that is essential for learning and validating surrogate constitutive equations in the context of machine learning models.

\begin{acknowledgments}
The authors acknowledge the financial support of the Lubrizol Corporation. N.K. also acknowledges conference grant support from the National University of Singapore Overseas Graduate Scholarship (NUS-OGS). 
\end{acknowledgments}

\section*{Data Availability Statement}
The data that support the findings of this study are available from the corresponding author upon reasonable request.
\appendix

\section{The Method of Dominant Balances for the Giesekus Model under Exponential Shear}

The Giesekus model \cite{giesekus_simple_1982} has a single nonlinear parameter $a$ that represents the degree of anisotropy in the drag force experienced by a polymer molecule modeled as a dumbbell, and is given by:
\begin{equation}
    \tau \accentset{\nabla}{\pmb{\sigma}} + \pmb{\sigma} + \frac{a \tau}{\eta_0} \pmb{\sigma} \cdot \pmb{\sigma} = \eta_0 \dot{\pmb{\gamma}}
\end{equation}

In a simple shear flow, we have the following set of nonlinear differential equations:
\begin{subequations}
\begin{align}
\begin{split} \label{giesekus11}
    \frac{d\sigma_{11}}{dt} + \frac{\sigma_{11}}{\tau} + (\sigma_{11}^2 + \sigma_{12}^2) \frac{a}{\eta_0} - 2\dot{\gamma} \sigma_{12} &= 0 
\end{split} \\
\begin{split} \label{giesekus22}
    \frac{d \sigma_{22}}{dt} + \frac{\sigma_{22}}{\tau} + (\sigma_{12}^2 + \sigma_{22}^2) \frac{a}{\eta_0} &= 0
\end{split} \\
\begin{split} \label{giesekus12}
    \frac{d \sigma_{12}}{dt} + \frac{\sigma_{12}}{\tau} + (\sigma_{11} + \sigma_{22}) \sigma_{12} \frac{a}{\eta_0} - \dot{\gamma} \sigma_{22} - \frac{\eta_0}{\tau} \dot{\gamma} &=0
\end{split} \\
\begin{split} \label{giesekus33}
    \frac{d \sigma_{33}}{dt} + \frac{\sigma_{33}}{\tau} + \sigma_{33}^2 \frac{a}{\eta_0} &= 0
\end{split}
\end{align}
\end{subequations}

Because the $\sigma_{33}$ term is factorizable from Eq. \ref{giesekus33}, $\sigma_{33}$ is set to trivially to be zero. However, in contrast to the linear PTT model, the evolution equation for $\sigma_{22}$ (Eq. \ref{giesekus22}) does not decouple and must be solved in conjunction with Eq. \ref{giesekus11} and \ref{giesekus12}.

For the case of an exponential shear strain rate function $\dot{\gamma}^e(t)$ (Eq. \ref{eq:expshear}(b)), we assume again that at long times, $\sigma_{12} \sim [\dot{\gamma}^e]^x$, $\sigma_{11} \sim [\dot{\gamma}^e]^y$ and $\sigma_{22} \sim [\dot{\gamma}^e]^z$. 

We start first with Eq. \ref{giesekus22}:
\begin{equation}
    z \alpha [\dot{\gamma}^e]^z + \frac{1}{\tau} [\dot{\gamma}^e]^z + ([\dot{\gamma}^e]^{2x} + [\dot{\gamma}^e]^{2z}) \frac{a}{\eta_0} = 0
\end{equation}
From a dominant balance of the nonlinear terms, we can conclude that $x = z$, but no more than that.

Moving on to Eq. \ref{giesekus11},
\begin{equation}
    y \alpha [\dot{\gamma}^e]^y + \frac{1}{\tau} [\dot{\gamma}^e]^y + ([\dot{\gamma}^e]^{2y} + [\dot{\gamma}^e]^{2x}) \frac{a}{\eta_0} - 2[\dot{\gamma}^e]^{1+x} = 0
\end{equation}

This balance is more tricky to identify. The nonlinear term must balance the applied strain rate, so either $2y = 1 +x$ or $2x = 1+x$. However, physically, at large strain rates, we expect that the normal stress should grow to be larger than the shear stress. Hence, the appropriate balance is $2y = 1 + x$. 

In Eq. \ref{giesekus12},
\begin{equation}
    x \alpha [\dot{\gamma}^e]^x + \frac{1}{\tau} [\dot{\gamma}^e]^x + ([\dot{\gamma}^e]^y + [\dot{\gamma}^e]^z) [\dot{\gamma}^e]^x \frac{a}{\eta_0} - [\dot{\gamma}^e]^{1+z} - \frac{\eta_0}{\tau} \dot{\gamma}^e = 0
\end{equation}

If we try to balance the nonlinear term with the strain rates, we get either $y + x = 1+z$ or $y+x = 1$. In the first case, this leads to $y= 1$, which then leads to $x = z = 1$. This is nonphysical because of our expectation that the normal stress must grow faster than the shear stress. In the second case, this leads to $y = 2/3$ and $x = z = 1/3$. This is inconsistent because the other strain rate term with the exponent $1 + z = 4/3$ is now larger than the assumed dominant balance which has an exponent 1. We are thus left to conclude that the strain rate terms must be the dominant balance, leading to $1+z = 1$. This means that $x = z = 0$ and $y = 1/2$, and this is in agreement with time-dependent simulations of Eq. \ref{giesekus11} to \ref{giesekus33} with an exponentially increasing shear rate given by Eq. \ref{eq:expshear}.

\begin{figure*}
\centering
\includegraphics[width=0.9\textwidth]{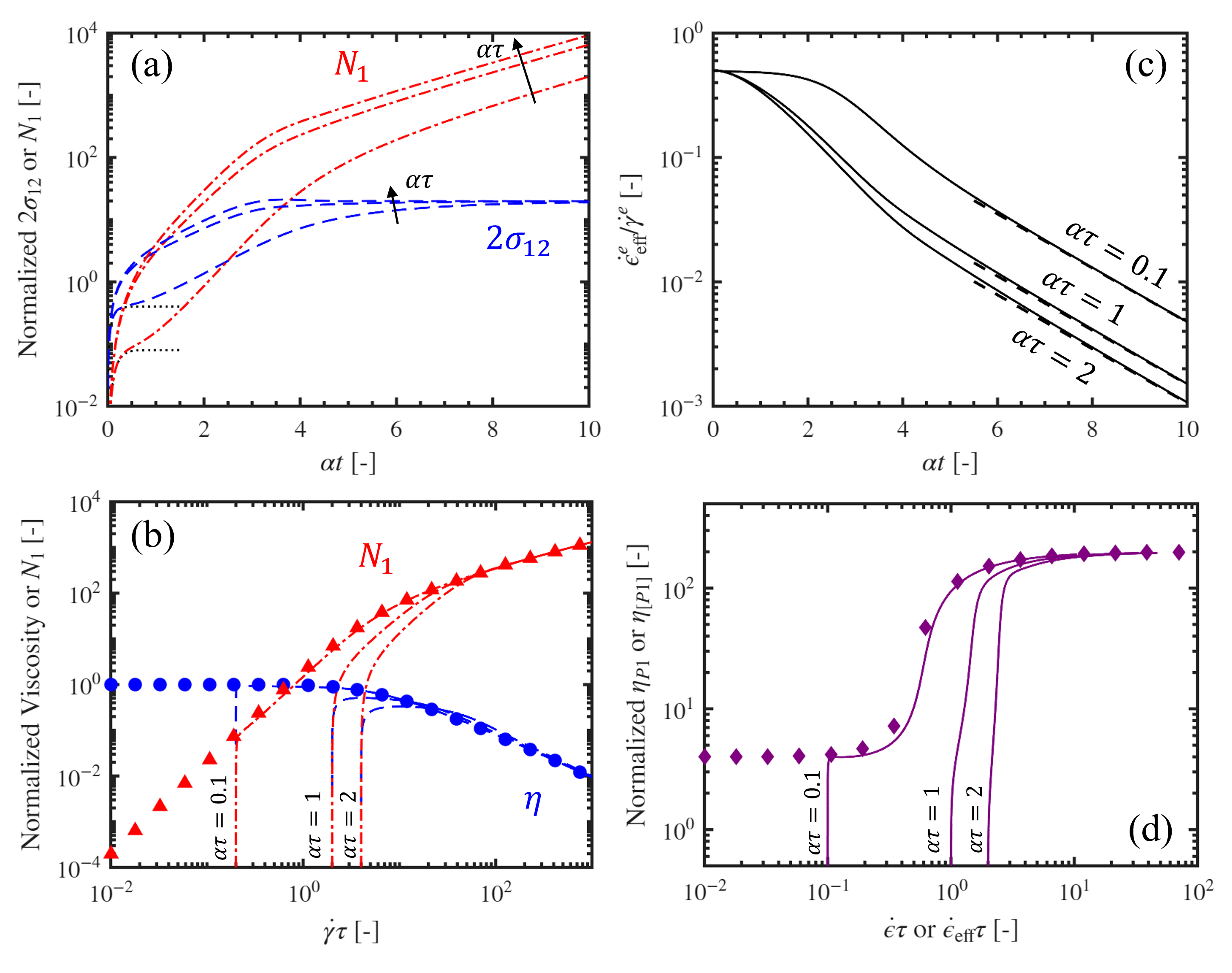}
\caption{Exponential shear flow of the Giesekus model with mobility parameter $a = 0.01$ for different dimensionless flow strengths ($\alpha \tau = 0.1, 1, 2$). All stress quantities are normalized by the modulus $G = \eta_0 / \tau$, viscosities are normalized by the shear viscosity $\eta_0$, and shear and extension rates by $1/\tau$. All symbols correspond to steady state data and lines show the trajectory of material responses computed from the time-evolving shear stress and $N_1$ measured in exponential shear flow. (a) The shear stress and first normal stress difference against dimensionless time. Curves of increasing $\alpha \tau$ are arranged from bottom to top. The leading order responses to an initial step strain rate for $\alpha \tau = 0.1$ are indicated by black dotted lines for both the shear stress and $N_1$. (b) The steady shear viscosity and first normal stress difference against a dimensionless strain rate. (c) Evolution of the normalized effective extension rate against dimensionless time. Curves of increasing $\alpha \tau$ are arranged from left to right. (d) The composite planar extensional viscosity function against effective extension rate, which all plateau to the same value at large extension rates. Curves of increasing $\alpha \tau$ are labeled from left to right. The symbols correspond to the true planar extensional viscosity computed analytically.} 
\label{fig:giesekus}
\end{figure*}

The surprising fact that $\sigma_{12}$ is constant with increasing strain rate (or large Weissenberg numbers) was explained in the following way by Khair \cite{khair_large_2016}: ``a physical explanation is that at large Wi the polymer chains are mostly aligned with the strong flow, which generates a constant shear stress.'' Furthermore, he goes on to say that the time derivatives of $\sigma_{11}$, $\sigma_{12}$ or $\sigma_{22}$ did not enter the leading order balances, indicating that ``relaxation (or memory) plays a secondary role compared to flow-driven orientation and anisotropic drag at large Wi; the stress evolves quasisteadily''. This idea of ``quasisteadiness'' was also invoked in Section II to explain the ability of exponential shear to obtain the steady shear and planar extension material functions.

The above results also match the scaling from steady shear as derived in Table 2 of Giesekus' original paper \cite{giesekus_simple_1982} as well, which was also reported in Table 7.3-5 of the Dynamics of Polymer Liquids textbook \cite{bird_dynamics_1987}. The scalings for $\dot{\gamma} \rightarrow \infty$ are:
\begin{subequations}
\begin{equation}
    \eta \sim \sqrt{\frac{1-a}{a}} \frac{\eta_0}{\tau \dot{\gamma}}
\end{equation}
\begin{equation}
    \Psi_1 \sim \frac{\sqrt{2}}{a} (a(1-a))^{1/4} \frac{\eta_0 \tau}{(\tau \dot{\gamma})^{3/2}}
\end{equation}
\end{subequations}

We show simulations of the Giesekus model subject to exponential shear in Fig. \ref{fig:giesekus}. Similar to the linear PTT model, the predictions using the material function Eq. \ref{expviscosities} match the steady state material functions at long times, as shown in Fig. \ref{fig:giesekus}(b). The effective extension rate normalized by instantaneous shear rate as a function of dimensionless time is shown in Fig. \ref{fig:giesekus}(c). This starts at a value of 0.5 and then slowly decreases with a power law of $-1/2$. This is because the value of $\dot{\epsilon}_\mathrm{eff}^e / \dot{\gamma}^e \rightarrow \sigma_{12} /N_1 = \sigma_{12} /(\Psi_1 \dot{\gamma}^2) \rightarrow (a(1-a)/4)^{1/4} (\tau \dot{\gamma})^{-1/2}$ at large strain rates.

The steady planar extensional viscosity for the Giesekus model has an analytical solution:
\begin{equation}
    \eta_{P1} (\dot{\epsilon}) = \frac{\eta_0}{2 a \tau \dot{\epsilon}} \left( \sqrt{(1 - 2\tau \dot{\epsilon})^2 + 8a\tau \dot{\epsilon}} - \sqrt{(1 + 2 \tau \dot{\epsilon})^2 - 8a \tau \dot{\epsilon}} + 4\tau \dot{\epsilon} \right)
\end{equation}
This material function eventually saturates at high extension rates to a value $\eta_P \rightarrow 2 \eta_0 / a$. For all values of $\alpha$ chosen, the predictions from Eq. \ref{compositematerialfunction} using exponential shear match the steady planar extensional viscosity curve at long times. However, to obtain accurate results for low extension rates as well, a low value of $\alpha$ (e.g. $\alpha \tau \approx 0.1$) should be chosen to fulfill the criteria $\alpha \tau \ll 1$.

\section{Rolie-Poly model}

\begin{figure*}
\centering
\includegraphics[width=0.9\textwidth]{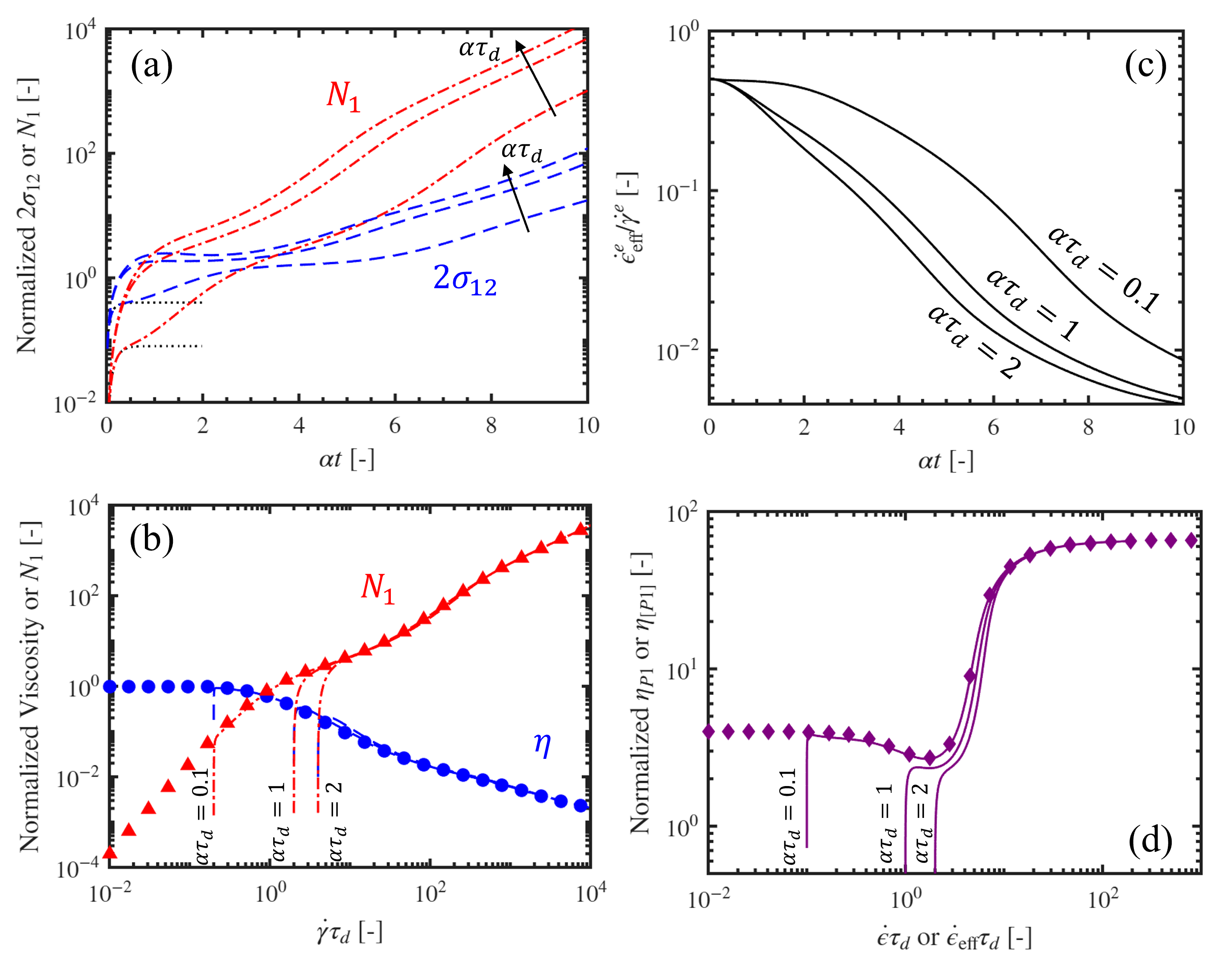}
\caption{Exponential shear flow of the Rolie Poly model with parameters $\tau_d / \tau_R = 5$, $\beta = 0.1$, $\delta = -0.5$, $L_{max} = 10$ for different dimensionless flow strengths ($\alpha \tau_d = 0.1, 1, 2$). All stress quantities are normalized by $G$, viscosities are normalized by the viscosity $G\tau_d$, and shear and extension rates by $1/\tau_d$. All symbols correspond to steady state data and lines show the trajectory of material responses computed from the time-evolving shear stress and $N_1$ measured in exponential shear flow. (a) The shear stress and first normal stress difference against dimensionless time. Curves of increasing $\alpha \tau_d$ are arranged from bottom to top. The leading order responses to an initial step strain rate for $\alpha \tau_d = 0.1$ are indicated by black dotted lines for both the shear stress and $N_1$. (b) The steady shear viscosity and first normal stress difference against a dimensionless strain rate. (c) Evolution of the normalized effective extension rate against dimensionless time. Curves of increasing $\alpha \tau_d$ are arranged from left to right. (d) The composite planar extensional viscosity function against effective extension rate, which all plateau to the same value at large extension rates. Curves of increasing $\alpha \tau_d$ are labeled from left to right. The symbols correspond to the true planar extensional viscosity computed analytically.} 
\label{fig:roliepoly}
\end{figure*}

The state of the art continuum model for entangled polymer melts is the Rolie-Poly model \cite{likhtman_simple_2003}. In terms of the polymer configuration tensor $\pmb{c}$ and the identity tensor $\pmb{I}$, the Rolie-Poly model is written as:
\begin{equation}
    \accentset{\nabla}{\pmb{c}} = - \frac{1}{\tau_d} (\pmb{c} - \pmb{I}) - \frac{2 k_s (\lambda)}{\tau_R} \left(1 - \frac{1}{\lambda}\right) \left(\pmb{c} + \beta \lambda^{2\delta} (\pmb{c} - \pmb{I}) \right)
\end{equation}
where the disengagement time is $\tau_d$, and the chain stretch $\lambda = \sqrt{\text{tr}~\pmb{c} / 3}$ evolves separately with the Rouse time $\tau_R$. Two scalar model parameters $\beta$ and $\delta$ modulate the effectiveness of convective constraint release and the chain sensitivity to stretching. The finite extensibility of the polymer chains is incorporated by applying the Cohen-Padé approximation to the inverse Langevin function for the spring law \cite{kabanemi_nonequilibrium_2009}, making the spring constant $k_s (\lambda)$ a function of the maximum extensibility ($L_{max}$) of the polymer chain:
\begin{equation}
    k_s(\lambda) = \frac{(3 - \lambda^2 / L_{max}^2)(1 - 1/ L_{max}^2)}{(1 - \lambda^2 / L_{max}^2)(3 - 1 / L_{max}^2)}
\end{equation}
The stress is related to the polymer configuration tensor $\pmb{c}$, the spring nonlinearity $k_s (\lambda)$ and the plateau modulus $G$:
\begin{equation}
    \pmb{\sigma} = G k_s (\lambda) (\pmb{c} - \pmb{I})
\end{equation}

When subject to exponential shear, we again see good agreement between the proposed material functions and the steady state material functions, as shown in Fig. \ref{fig:roliepoly}. Even though the ``method of dominant balances'' is more difficult to explicitly apply here, the underlying idea still holds. Notably, our new proposed material function Eq. \ref{compositematerialfunction} is able to replicate the planar extensional viscosity curve, with the initial dip at $\dot{\epsilon} \tau_d \simeq 1/2$ due to chain orientations within the tube, and later the sharp increase at a critical Weissenberg number (now defined in terms of the Rouse time) corresponding to $\dot{\epsilon} \tau_R \simeq 1$, or equivalently $Wi_{CS} \equiv \dot{\epsilon}_{CS} \tau_d \simeq \tau_d / \tau_R$, and finally a saturation due to finite extensibility effects.

\section{Cox-Merz and Laun's Rule for Experimental Data}

We can also assess whether both polymeric fluids used in this study follow the Cox-Merz rule \cite{cox_correlation_1958} and Laun's rule \cite{laun_prediction_1986}. If they do follow these rules, we can potentially use the linear viscoelastic measurements directly without a need for the quasisteady measurements of the viscosity and first normal stress difference over a range of deformation rates from exponential shear. The Cox-Merz rule \cite{cox_correlation_1958} is an empirical relationship which states that the nonlinear shear viscosity curve can be approximated by the magnitude of the complex viscosity obtained from linear viscoelasticity measurements as a function of angular frequency:
\begin{equation}
    \eta(\dot{\gamma})|_{\dot{\gamma}=\omega} = |\eta^*(\omega)| = \frac{G''(\omega)}{\omega} \left[1 + \left(\frac{G'(\omega)}{G''(\omega)} \right)^2 \right]^{0.5}
\end{equation}

Closely related is Laun's rule \cite{laun_prediction_1986}, first proposed for low-density polyethylene melts, which also posits an empirical relationship between the nonlinear first normal stress coefficient and the magnitude of the complex viscosity:
\begin{equation}
    \Psi_1(\dot{\gamma})|_{\dot{\gamma}=\omega} = \frac{2 G'(\omega)}{\omega^2} \left[1 + \left(\frac{G'(\omega)}{G''(\omega)} \right)^2 \right]^\ell
\end{equation}

Laun \cite{laun_prediction_1986} originally suggested that $\ell = 0.7$, but Das \textit{et al.} \cite{das_launs_2024} later showed how a more optimal value of $\ell$ can be found by applying an extended theoretical framework that combines linear viscoelastic data, a fit to a fractional Maxwell liquid model, and nonlinear step strain data. In the absence of step strain data, we use $\ell = 0.7$ for the data here.

\begin{figure*}
\centering
\includegraphics[width=\textwidth]{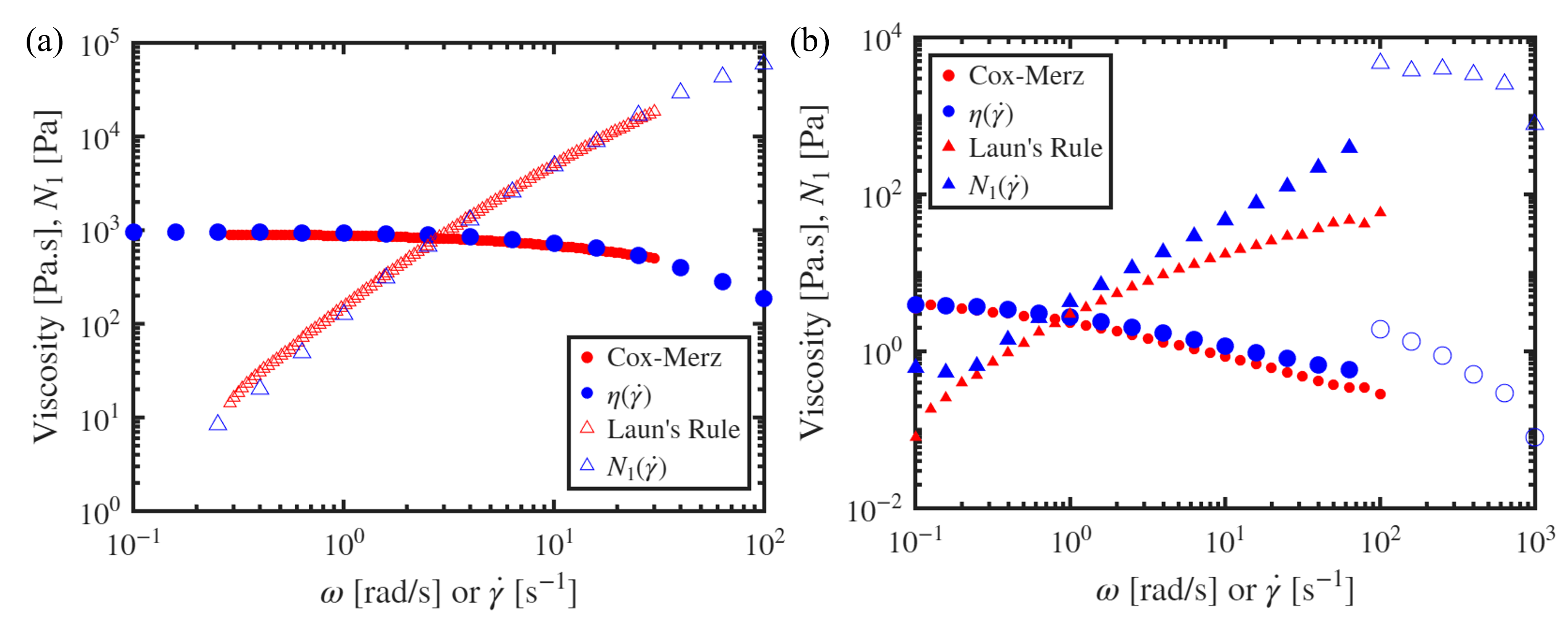}
\caption{Cox-Merz and Laun's rule predictions compared with steady shear data for (a) PDMS which has a zero-shear viscosity $\eta_0 = 897$ Pa.s and viscosity-weighted relaxation time $\bar{\tau}_\eta = 0.107$ s, and (b) the PIB solution, which has $\eta_0 \approx 5$ Pa.s and $\bar{\tau}_\eta \approx 2.92$ s. Larger symbols are the steady shear data and smaller symbols are from the empirical rules. Circle symbols represent viscosity and triangle symbols represent $N_1$.} 
\label{fig:coxmerz}
\end{figure*}

As shown in Fig. \ref{fig:coxmerz}(a), the well-entangled PDMS melt follows both the Cox-Merz rule and Laun's rule evaluated using $\ell = 0.7$ surprisingly well. On the other hand, the PIB solution (Fig. \ref{fig:coxmerz}(b)) follows the Cox-Merz rule reasonably well, but not Laun's rule. We hence expect that measurements of the SAOS behavior of the PDMS sample over a greater range of frequencies can lead to reliable extrapolation of its nonlinear shear thinning behavior to even higher strain rates beyond the rates experimentally accessible using steady shear (i.e. not impeded by edge fracture). Similarly, the use of Laun's rule can extend knowledge of the expected variation of $N_1(\dot\gamma)$. These two functions can then be used in Eqs. \ref{expviscosities} and \ref{compositematerialfunction} to reliably extrapolate the expected variation in $\eta_{[P1]}(\dot{\epsilon}_\mathrm{eff}^e)$ to higher Weissenberg numbers.

% \nocite{*}
\bibliographystyle{ieeetr}
\bibliography{ExponentialShear}% Produces the bibliography via BibTeX.
% \bibliography{aipsamp}% Produces the bibliography via BibTeX.

\end{document}